\newif\ifdraft
\drafttrue

\documentclass[compsoc]{IEEEtran}

\usepackage{url}

\usepackage{cite}
\usepackage{nameref}	
\usepackage{array}

\usepackage{xr}

\usepackage[american]{babel}
\usepackage{graphicx}
\usepackage{subcaption}
\usepackage{caption}
\usepackage{multirow}

\usepackage{siunitx}
\DeclareSIUnit{\molar}{M}

\usepackage[dvipsnames]{xcolor}
\usepackage{pgfplots}
\usepackage{amsmath,amssymb,amsthm,amsfonts}
\usepackage{hyperref}
\usepackage{cleveref}
\usepackage{booktabs}
\usepackage{xparse}
\NewDocumentCommand{\DIV}{om}{%
	\IfValueT{#1}{\setcounter{#2}{\numexpr#1-1\relax}}%
	\csname #2\endcsname
}
\graphicspath{{./src/}}

\usepackage[normalem]{ulem} 
\newcommand{\stkout}[1]{\ifmmode\text{\sout{\ensuremath{#1}}}\else\sout{#1}\fi}

\ifdraft

\newcommand{\deleted}[1]{\textcolor{red}{\stkout{#1}}}

\newcommand{\deletedfloat}[1]{}
\else

\newcommand{\deleted}[1]{}

\newcommand{\deletedfloat}[1]{}
\fi

\newcommand{\gettitle}{Self-supervised representation learning from 12-lead ECG data}

\newcolumntype{L}[1]{>{\raggedright\let\newline\\\arraybackslash\hspace{0pt}}m{#1}}
\newcolumntype{C}[1]{>{\centering\let\newline\\\arraybackslash\hspace{0pt}}m{#1}}
\newcolumntype{R}[1]{>{\raggedleft\let\newline\\\arraybackslash\hspace{0pt}}m{#1}}

\hypersetup{
    pdftitle={\gettitle},
    pdfauthor={Temesgen Mehari and Nils Strodthoff},
    pdfkeywords={electrocardiography} {deep neural networks}
    bookmarksopen=false,
    bookmarksnumbered=true
}

\begin{document}

\title{\gettitle}

\author{Temesgen Mehari \& Nils Strodthoff \thanks{Temesgen Mehari is with Physikalisch Technische Bundesanstalt, Berlin, Germany and Fraunhofer Heinrich Hertz Institute, Berlin, Germany, email: temesgen.mehari@ptb.de. Nils Strodthoff is with Fraunhofer Heinrich Hertz Institute, Berlin, Germany, e-mail: nils.strodthoff@hhi.fraunhofer.de. Corresponding author: Nils Strodthoff (Permanent address: University of Oldenburg, Germany, e-mail: nils.strodthoff@uol.de). Both authors contributed equally to this work.}}

\maketitle
\begin{abstract}
Clinical 12-lead electrocardiography (ECG) is one of the most widely encountered kinds of biosignals. Despite the increased availability of public ECG datasets, label scarcity remains a central challenge in the field. Self-supervised learning represents a promising way to alleviate this issue. This would allow to train more powerful models given the same amount of labeled data and to incorporate or improve predictions about rare diseases, for which training datasets are inherently limited. In this work, we put forward the first comprehensive assessment of self-supervised representation learning from clinical 12-lead ECG data. To this end, we adapt state-of-the-art self-supervised methods based on instance discrimination and latent forecasting to the ECG domain. In a first step, we learn contrastive representations and evaluate their quality based on linear evaluation performance on a recently established, comprehensive, clinical ECG classification task. In a second step, we analyze the impact of self-supervised pretraining on finetuned ECG classifiers as compared to purely supervised performance.
For the best-performing method, an adaptation of contrastive predictive coding, we find a linear evaluation performance only 0.5\% below supervised performance. For the finetuned models, we find improvements in downstream performance of roughly 1\% compared to supervised performance, label efficiency, as well as robustness against physiological noise. This work clearly establishes the feasibility of extracting discriminative representations from ECG data via self-supervised learning and the numerous advantages when finetuning such representations on downstream tasks as compared to purely supervised training. As first comprehensive assessment of its kind in the ECG domain carried out exclusively on publicly available datasets, we hope to establish a first step towards reproducible progress in the rapidly evolving field of representation learning for biosignals.

\end{abstract}

\IEEEpeerreviewmaketitle

	\section{Introduction}
The availability of datasets with high-quality labels is an omnipresent challenge in machine learning in general, but especially in the health domain, where the labeling process is particularly expensive and clinical ground truth is in many cases hard to define. However, the amount of unlabeled data often exceeds the amount of labeled data by several orders of magnitude, which represents a strong case for (self-supervised) representation learning from unlabeled data. During the past few years, self-supervised learning has made enormous advances in different domains ranging from natural language processing \cite{devlin-etal-2019-bert} over speech \cite{oord2018representation} to computer vision \cite{chen2020simple}. Self-supervised learning could be one component towards addressing the problem of data scarcity. It could help to train more accurate and potentially also more robust models given the same amount of labeled data, which is a desirable prospect for any application field. Of particular importance for the medical domain are improvements in label efficiency, which could allow to train models on more finegrained and consequently less populated label hierarchies, or to include rare diseases that were out of reach with conventional training methods.

In this work, we investigate self-supervised representation learning in the context of clinical electrocardiography (ECG) data. The ECG is a non-invasive method that allows to assess the general cardiac condition of a patient. It is therefore an important tool for the first-in-line examination for the diagnosis of cardiovascular diseases, which rank among the diseases of highest mortality \cite{wilkins2017statistics}. In particular, the (short) 12-lead ECG, which we focus on in this work, is the most commonly used type of ECG with a very broad clinical applicability ranging from primary care centers to intensive care units. Even though the technology underlying the ECG is by now more than 100 years old and it is an extremely common procedure, which is ordered or provided during 5\% of the office visits in the US \cite{NACMS2016}, its interpretation is still performed mostly manually with only limited algorithmic support. Here, it is important to recognize that ECG interpretation is in some cases even challenging for cardiologists \cite{Salerno2003}. 

There are deep-learning-based ECG interpretation algorithms with exceptionally high performance \cite{Kashou2020,Ribeiro2020} that have been trained on large closed-source datasets. The sizes of publicly available datasets are smaller by several orders of magnitude, which provides the motivation to investigate if and how far self-supervised learning techniques can improve the performance of algorithms trained on these datasets. In addition, the question of label quality remains challenging even for the above large-scale datasets. Here, it is important to stress that even though self-supervised methods have been applied successfully in computer vision and speech, ECG records are timeseries (rather than a one-dimensional image) and multivariate data (unlike speech) with considerably different properties than speech. This means that the degree to which self-supervised methods work in this domain is not clear from the onset and deserves a thorough study. Even though the underlying methods are based on methods developed for other application domains, it requires subtle adaptations, such as a careful choice of augmentation transformations or adaptations of the model architecture and training procedure, to make them actually work in the context of ECG data. And finally, just as in the case of supervised learning \cite{Strodthoff:2020Deep}, measurable progress in the field of representation learning for ECG data requires benchmarking based on clearly defined evaluation criteria, in the ideal case with open software on publicly accessible datasets. With this work, we hope to establish a first step in this direction. 

Going beyond a mere benchmarking of representation learning algorithms for the ECG domain, the real benefits lie in the potential benefits of self-supervised pretraining for finetuned downstream classifiers. This includes aspects such as including improved data efficiency, improved quantitative performance, or improved robustness in a general sense
as compared to model trained in a purely supervised fashion. In our experimental results, we present explicit evidence for these benefits. Putting these results into perspective, demonstrating significant improvement through self-supervised pretraining in terms of downstream performance should not be taken for granted as the effects often remain small \cite{azizi2021big}. Also improved robustness properties from self-supervised have rarely been demonstrated explicitly in other domains not to mention the domain of ECG data. 

Our key achievements can be summarized as follows:
\begin{itemize}
	\item We present the first comprehensive assessment of self-supervised representation learning for 12-lead ECG data to foster measurable progress in the subfield of representation learning for biosignals.
	\item We adapt and directly compare instance-based self-supervised methods (Simple Framework for Contrastive Learning of Visual Representations \textit{(SimCLR)}, Bootstrap Your Own Latent \textit{(BYOL)}, Swapping Assignments between multiple Views \textit{(SWaV)}) and contrastive, latent forecasting methods (contrastive predictive coding \textit{(CPC)}) and find compelling evidence for the feasibility of learning useful representation from ECG data through self-supervised learning.
	\item We propose and evaluate several modifications in the CPC architecture and training procedure that lead to considerable performance improvements. 
	\item We evaluate different quality aspects of downstream classifiers finetuned from self-supervised models compared to training from scratch and find evidence for improved quantitative performance given the same downstream training set, improved label efficiency and improved robustness through self-supervised pretraining.
\end{itemize}

	\begin{figure*}[ht]
	\centering
	\includegraphics[width=.95\textwidth]{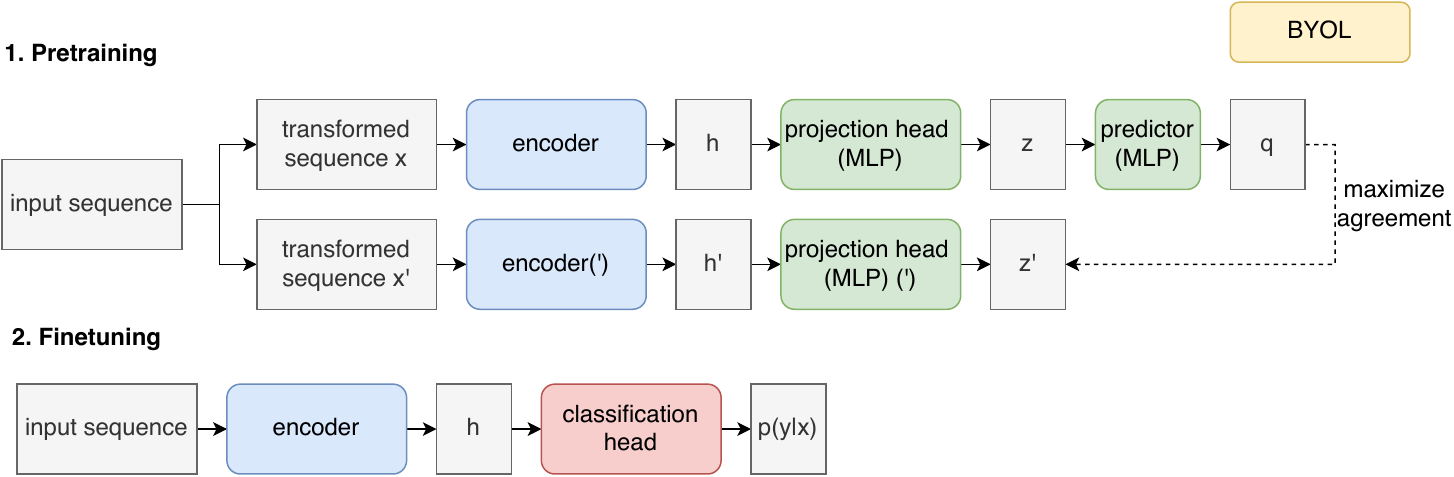}
	\caption{Schematic illustration of contrastive methods based on instance-discrimination for the case of BYOL.}
	\label{fig:schematic_simclr}
\end{figure*}

\section{Materials and methods}
\subsection{ECG analysis using deep learning}
By now, the analysis of ECG data has developed into a very popular application domain for deep learning methods. As we are solely focusing on deep learning methods in this work, a brief discussion of state-of-the-art methods in the field is appropriate. For a detailed comparison of different approaches, we refer the reader to recent reviews \cite{Hong2020,Minchol2019}. Until very recently, the merits of newly proposed methods in the field have been very difficult to assess due to a lack of appropriate, large, publicly available ECG datasets for training and evaluation of algorithms as well as due to a lack of clearly defined benchmarking criteria. The first issue was resolved very recently with the publication of several large clinical ECG datasets, see \Cref{sec:datasets} for details. Relating to the second issue, we draw on a recent benchmarking study \cite{Strodthoff:2020Deep} on the \textit{PTB-XL} dataset \cite{Wagner:2020PTBXL}, the very dataset also used as downstream dataset in this study, where a number of different classification algorithms was assessed on different clinically relevant ECG classification tasks. The overall best-performing methods in this study turned out to be modern convolutional network architectures, namely resnet- or inception-based architectures. This is in line with the network architectures used in the literature \cite{Hannun2019,Ribeiro2020,Kashou2020} for which the authors report excellent performance on datasets with restricted access that are up to several orders of magnitude larger than the currently available public datasets.

\subsection{Self-supervised representation learning for ECG data}
Contrastive methods in computer vision have witnessed tremendous advances in the past few months \cite{he2020momentum,chen2020simple,chen2020improved,grill2020bootstrap,caron2021unsupervised}, which significantly improved the linear evaluation performance on ImageNet and demonstrated the usefulness of the learned features for other computer vision tasks. These methods can be adapted straightforwardly to learning representations from a large number of relatively short time series segments if one interprets the time series record as a one-dimensional multichannel image and adapts transformations appropriate for time series. The predominantly used approaches rely on instance discrimination as pretraining task and will be discussed in \Cref{sec:instance} in detail. A second domain, where self-supervised methods for non-discrete data have been implemented successfully is the domain of representation learning speech, where predictive coding methods \cite{oord2018representation,chung2019unsupervised} have been applied to conventional acoustic features \cite{chung2019unsupervised,blandon2020analysis,vanstaden2020comparison} but also to raw waveform data \cite{oord2018representation,schneider2019wav2vec,baevski2020wav2vec}. The best-performing methods in this field rely on latent forecasting tasks and are discussed in \Cref{sec:cpc}.

\subsubsection{Instance discrimination (\textit{SimCLR/BYOL/SwAV})}
\label{sec:instance}
Current state-of-the-art contrastive methods from computer vision aim to learn representations based on multiple views on the same instance, see \Cref{fig:schematic_simclr}. These are created by applying stochastic transformations to the input data. This idea is implemented in the most straightforward way in \textit{SimCLR} \cite{chen2020simple}, where a noise contrastive loss is used to attract two (positive) copies originating from the same original instance and to repel instances from all other (negative) instances in the batch, an approach which typically relies on training with large batch sizes, which is less problematic in our case due to the reduced dimensionality of time series data as compared to image data. \textit{BYOL} \cite{grill2020bootstrap} does not explicitly rely on contrasting against negative samples in the same batch, but uses a moving average of the model itself and reported slight improvements over \textit{SimCLR} in the image domain. Finally, \textit{SwAV} \cite{caron2021unsupervised} relies on contrasting cluster assignments rather than individual instances and once again improved the linear evaluation scores on ImageNet. In our case, we build on the implementations of all three frameworks in \textit{PyTorch Lightning Bolts} \cite{falcon2020framework}.

As model architecture, we use the convolutional neural networks of the \textit{xresnet1d}-family, one-dimensional adaptations of the popular \textit{xresnet}-architecture \cite{he2019bag} from computer vision, which showed very good perform in a very recent ECG classification benchmarking study \cite{Strodthoff:2020Deep} that was carried out on the \textit{PTB-XL} dataset, see \Cref{sec:datasets}, using the same evaluation scheme that is supposed to be used in this work, see \Cref{sec:traineval}. We base our experiments on a \textit{xresnet1d50}, whose performance is compatible with that of the best-performing \textit{xresnet1d101} from \cite{Strodthoff:2020Deep} within error bars. However, it is more parameter-efficient and showed a slightly superior performance for linear evaluation and finetuning. At this point, we stress again that even though we use an architecture that achieves state-of-the-art performance on \textit{PTB-XL}, the main point of our study lies in the demonstration of relative improvements compared to supervised performance. 

The transformations used to generate two semantically equivalent views of a given original record lie at the heart of the recent success of contrastive methods in computer vision. As demonstrated in \cite{chen2020simple}, the quality of the learned representations depends crucially on the choice and proper combination of transformations. We therefore evaluated a number of transformations inspired by effective transformations in computer vision and transformations specific for time series, see \Cref{sec:transformations} for a detailed description. Finally, we also evaluate representations obtained by using only prototypical physiological noise during pretraining.

\begin{figure*}[ht]
	\centering
	\includegraphics[width=.95\textwidth]{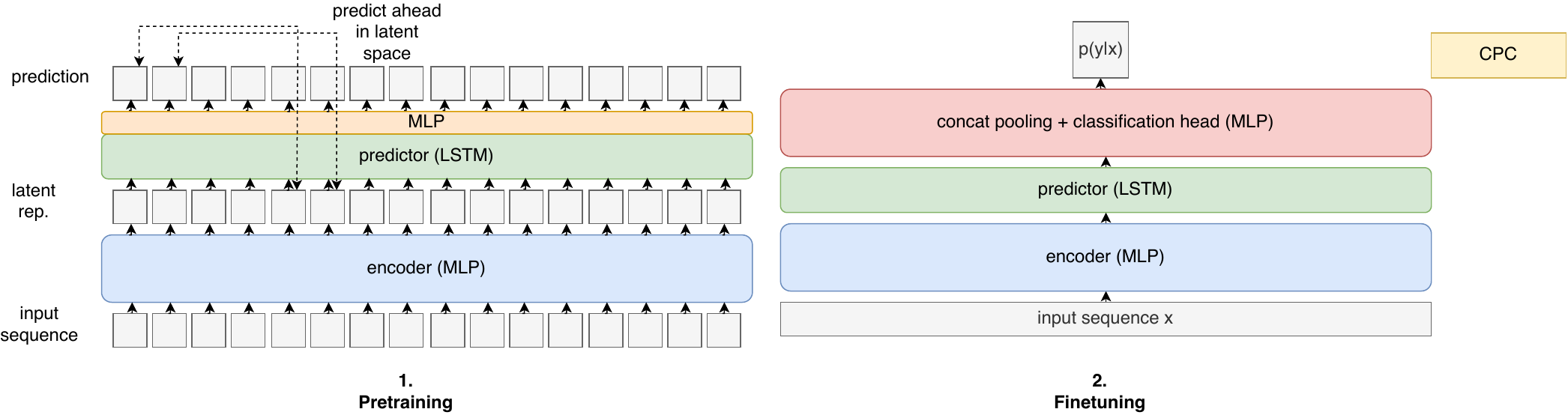}
	\caption{Schematic illustration of pretraining and finetuning procedure for contrastive predictive coding (\textit{CPC}).}
	\label{fig:schematic_cpc}
\end{figure*}

\subsubsection{Latent forecasting (\textit{CPC})}
\label{sec:cpc}
Contrastive Predictive Coding (\textit{CPC}) \cite{oord2018representation} is also a contrastive approach, which, in contradistinction to the approaches described above, explicitly makes use of the sequential ordering of the data. The idea is to encode the input sequence by means of an encoder with strided convolutions or fully connected layers and train a model to predict the latent representation of the sequence a fixed number of steps in the future given the encoded representation of the sequence in the past again using a noise contrastive estimation approach, see \Cref{fig:schematic_cpc} for a graphical representation. As we work with data at 100Hz, which is sampled rather coarsely compared the typical sampling rates of 10~kHz in the audio domain, there is no need to drastically downsample the signal by means of strided convolutions. Instead, we use a fully connected encoder, in our case composed of four layers with 512 filters with batch normalization, as it was also done in self-supervised representation learning from classical audio features \cite{chung2019unsupervised,blandon2020analysis,vanstaden2020comparison}. We predict 12 timesteps or equivalently to 0.12s into the future and work with 128 false negatives that are drawn from the same sequence as the original record. For the prediction task, we use a LSTM model \cite{hochreiter1997long} with 2 layers and 512 hidden units.  We propose and evaluate an enhanced version of the \textit{CPC} architecture, with an additional hidden layer and non-linearity before the linear output layer of the LSTM. This modification was inspired by the additional multilayer perceptron in \textit{SimCLR}, which was one of the key components that lead to superior performance compared to previously used self-supervised approaches in computer vision.

When finetuning a classification model, we apply a concat-pooling layer \cite{howard2018universal}, which concatenates the maximum of all LSTM outputs, the mean of all LSTM outputs, and the LSTM output corresponding to the final step, and a fully connected classification head with a single hidden layer with 512 units including batch normalization and dropout for regularization. To assess the linear evaluation performance, we use a single fully connected layer on top of the concat-pooling layer. The effect of the different modifications compared to standard \textit{CPC} implementations and finetuning schedules are investigated in detail in \Cref{sec:cpc_ablations}.

\subsubsection{Self-supervised representation learning for physiological time series data}
Self-supervised methods have also been used for representation learning from biomedical sequence data, including, most prominently, ECG \cite{Yuan2019,cheng2020subjectaware,Sarkar2020,kiyasseh2020clocs} and electroencephalography (EEG) \cite{Yuan2019,banville2019selfsupervised,cheng2020subjectaware,banville2020uncovering} data. With the exception of \cite{kiyasseh2020clocs}, none of the existing works considered the case of representation learning from clinical 12-lead ECGs, the clinically most widely encountered type of ECG measurement. The authors of \cite{kiyasseh2020clocs} also consider BYOL and \textit{SimCLR} for pretraining but used a very shallow network architecture with only five layers. We believe that it is necessary to use larger models, which reach state-of-the-art performance on large, comprehensive ECG datasets such as \emph{PTB-XL} and which consequently allow to learn richer representations, along with pretraining on larger datasets. In particular, it is not clear if pretraining advances in the small model regime carry over to larger models with competitive supervised performance. Also, from the methodological point of view, \cite{kiyasseh2020clocs} deviates significantly from our approach in the sense that they aim to learn a lead-independent, universal single-lead encoder, as opposed to a joint 12-lead encoder in our case. The former is not directly applicable to downstream 12-lead ECG tasks.In addition, they propose new contrastive methods that can use 12-lead ECG data during pretraining but differ from our methods in that they are not expedient for downstream 12-lead ECG tasks. This is because the proposed models do not process 12-lead data directly but exploit the fact that different leads from the same patient during pretraining can be considered as positive pairs. 
From the methodological point of view, \cite{cheng2020subjectaware} is also close to our contrastive approach but their experimental results were limited to a small 2-lead dataset with less than 50 records. Without access to the original implementation, it is impossible to assess if their proposed approach would be competitive on 12-lead data and on large (pretraining) datasets, where self-supervised methods reveal their full potential. Earlier approaches such as \cite{Yuan2019} trained representations from 2-lead ECGs using skip-gram models. Finally, \cite{Sarkar2020} use transformation recognition as a pretext task and proposed a framework specific to representation learning from single-lead ECGs. 

\subsection{ECG datasets}
\label{sec:datasets}
We use a collection of three datasets for pretraining henceforth referred to as \textit{All}, namely \textit{CinC2020} \cite{CinC2020}, \textit{Ribeiro} \cite{Ribeiro2020} and \textit{Chapman} \cite{Zheng2020}, which constitutes a collection of the largest publicly available 12-lead ECG datasets with in total 54,566 records. It is worth mentioning that \textit{CinC2020}, the training dataset used for the Computing in Cardiology Challenge 2020, is by itself a compilation of five different datasets. In particular, it includes the \textit{PTB-XL} dataset \cite{Wagner:2020PTBXL,Goldberger2020:physionet} that we also use for evaluation in this study. At the most finegrained level, which is used here, the \textit{PTB-XL} dataset comes with 71 labels and the evaluation task is framed as a multi-label classification task. It is worthwhile stressing that these labels cover a wide variety of diagnostic, form and rhythm statements and can be used for a comprehensive evaluation of ECG analysis algorithms. The 44 diagnostic statements can be categorized in terms of five super classes (normal/conduction disturbance/myocardial infarction/hypertrophy/ST-T change), the 19 form statements relate to mostly morphological changes in specific ECG segments such as an abnormal QRS complex, and the 12 rhythm statements comprise statements characterizing normal cardiac rhythms as well as arrhythmia. The dataset is organized into ten stratified, label-balanced folds, where the first eight are used as training set, the ninth is used as validation set and the tenth fold serves as test set \cite{Wagner:2020PTBXL}. All datasets are summarized in \Cref{tab:datasets}.

\begin{table}[ht]
	\centering
	\caption{Dataset summary: For pretraining we use \textit{All} (\textit{CinC2020} and \textit{Chapman} and \textit{Ribeiro}) or \textit{PTB-XL}. We evaluate on \textit{PTB-XL}. Note that \textit{PTB-XL} is a subset of \textit{CinC2020}.}
    \label{tab:datasets}
    \begin{tabular}{l|rr}
    \toprule
	dataset& \#samples&\# patients\\\midrule
	\textbf{Pretraining:} \textit{All} & 54,566 & unknown \\\midrule
	-\textit{CinC2020} \cite{CinC2020} & 43,093 &  unknown\\

	-\textit{Chapman} \cite{Zheng2020} &  10,646 & 10,646 \\
	-\textit{Ribeiro} \cite{Ribeiro2020} & 827 & 827 \\\midrule\midrule
	\textbf{Evaluation} & 21,837 & 18,885\\\midrule
	- \textit{PTB-XL} \cite{Wagner:2020PTBXL} & 21,837 & 18,885 \\\midrule
\bottomrule
    \end{tabular}
\end{table}

\subsection{Training and evaluation protocol}
\label{sec:traineval}
We restrict ourselves to ECG data at a sampling rate of 100Hz in all cases. We pretrain CPC models on input sequences with a length of 10 seconds, all other models (including finetuned CPC models) are trained on input sequences with length of 2.5 seconds. During training, subsequences are randomly cropped from the input record. During test time while finetuning, we use test-time-augmentation and crop all sequences to a length 2.5 seconds (using a stride of 1.25 seconds) from the original record and take the mean of their respective output probabilities as final prediction, a method which considerably improved the model performance by approximately 0.01 in macro AUC as compared to a naive evaluation \cite{Strodthoff:2020Deep}. Both for finetuning and pretraining, we use the AdamW optimizer \cite{loshchilov2019decoupled} with a weight decay of 0.001. During pretraining, we optimize the InfoNCE loss \cite{oord2018representation} using a constant learning rate schedule for CPC and the respective contrastive losses with a cosine learning rate schedule for \textit{SimCLR}, \textit{BYOL}, and \textit{SwAV} as described in the original publications. During finetuning, we optimize binary crossentropy with a constant learning rate schedule as appropriate for a multi-label classification task and evaluate the model performance based on macro AUC as in \cite{Strodthoff:2020Deep}, computed from the 71 labels on the most finegrained level in \textit{PTB-XL} \cite{Wagner:2020PTBXL}.
We perform model selection on the validation and select the model with the lowest validation loss during pretraining and highest macro AUC during finetuning. We report the respective test set score of the selected best model. The source code to reproduce all our experiments is publicly available \cite{implementation}.\\
As conventionally done in self-supervised representation learning studies, we use two different evaluation procedures, \textit{linear evaluation} and \textit{finetuning}. The \textit{linear evaluation} protocol aims to assess the quality of the learned representations through the linear separability of the learned representations. To this end, we replace the classification head by a single linear layer and freeze all other layers as well as batch normalization statistics. Within the \textit{finetuning} protocol, we investigate the usefulness of these representations for downstream tasks, where we unfreeze the classification head as well as all layers of the pretrained model. For \textit{CPC}, we found it beneficial to follow a two-step approach during finetuning: In a first step, we finetune just the classification head for 50 epochs while keeping the remaining pretrained weights fixed but still updating batch normalization statistics. We perform model selection using validation set scores and then finetune the entire model for 20 epochs at a reduced learning rate using discriminative, i.e.\ layer-dependent learning rates to mitigate the danger of overwriting the information captured during pretraining, where we typically divide models into head, body and stem/encoder and reduce the learning rate by a factor of 10 compared to the respective previous layer group. Also in this case, we select the final model based on validation set scores. In all other cases, we train models for 100 epochs using a constant learning rate.

	\section{Experiments}
The performance of self-supervised contrastive methods based on instance discrimination crucially depends on the choice of transformations that are used to create two semantically equivalent copies of the original input sequence. To determine appropriate transformations, we carried out an experiment, where we pretrained a model using \textit{SimCLR} and different combinations of transformations and evaluated the quality of the learned representations based on linear evaluation performance on \textit{PTB-XL}, see \Cref{sec:model_selection} in the supplementary material for details. The results clearly identify \textit{time out (TO)} in combination with \textit{random resized crop (RRC)} as most effective transformation pair, see \Cref{sec:transformations} in the supplementary material for a description of all transformations under consideration. In addition, we consider physiological noise transformations that were designed to mimic typical physiological noise that might occur during ECG measurements, namely \textit{baseline wander}, \textit{powerline noise}, \textit{electromyographic noise} and \textit{baseline shift}. In a second step, we used a similar protocol to compare the three different contrastive learning frameworks \textit{SimCLR}, \textit{BYOL} and \textit{SwAV} now using the predetermined set of transformations. Whereas \textit{SimCLR} reached clearly the best linear evaluation performance, finetuning from a \textit{BYOL} representation lead to a superior downstream performance after finetuning, which is why we consider both methods in the following sections.

\subsection{Self-supervised pretraining learns meaningful representations from ECG data}
\label{sec:lineareval}
\begin{figure}[ht]
	\includegraphics[width=\columnwidth]{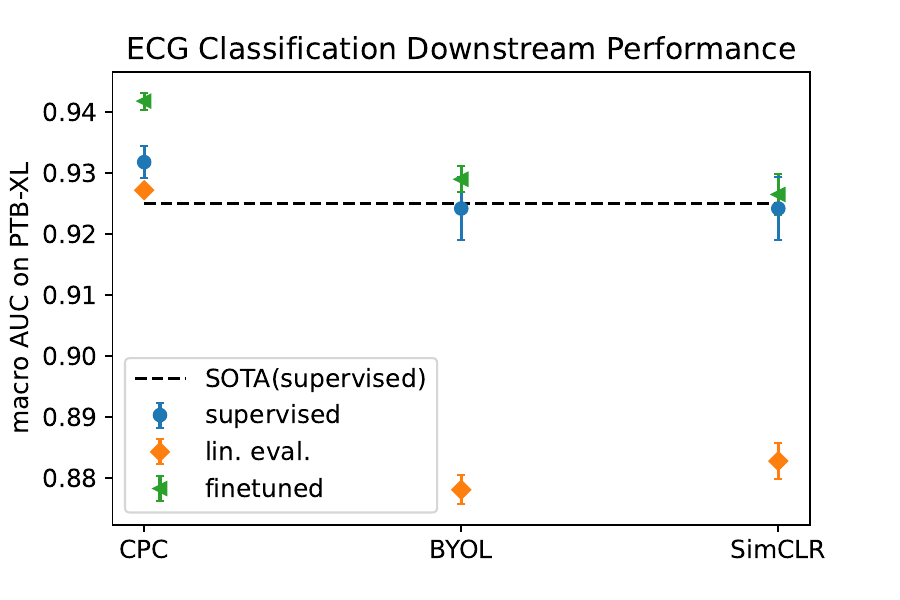}
	\caption{Comparison of different contrastive learning frameworks in terms of downstream performance comparing the three self-supervised learning frameworks \textit{CPC}, \textit{BYOL} and \textit{SimCLR}. The previous supervised state-of-the-art result from \cite{Strodthoff:2020Deep} is represented by a dashed line.}
	\label{fig:summary}
\end{figure}

\begin{table*}[t]
	\caption{Linear evaluation and finetuning performance on a comprehensive clinical downstream ECG classification task (macro AUC on the \textit{PTB-XL} test set). As before, we report mean and standard deviation over 10 finetuning runs.}
    \label{tab:summary}
    \centering
    \begin{tabular}{ll|ll}
    \toprule
    method & model& \multicolumn{2}{c}{\textit{PTB-XL}}\\

	&  &lin. eval. & finetuned \\
	
				\midrule
supervised  & 4FC+2LSTM+2FC & 0.7110(65) & 0.9318(26)\\
supervised  & xresnet1d50& 0.7210(158) &0.9242(51) \\\midrule\midrule
CPC (on All)&  4FC+2LSTM+2FC &\textbf{0.9272(08)}& \textbf{0.9418(14)} \\
CPC (on PTB-XL) &  4FC+2LSTM+2FC &  0.9246(08) &   0.9395(11)\\
CPC (on CinC2020 w/o PTB-XL) &  4FC+2LSTM+2FC & 0.9192(11)  & 0.9401(16)\\\midrule
SimCLR (RRC, TO)  & xresnet1d50&0.8828(29) & 0.9265(33)\\
SimCLR physio.  & xresnet1d50& 0.7701(31) &0.9258(13) \\
BYOL (RRC,TO) & xresnet1d50& 0.8781(24)&  0.9290(21)\\
BYOL  physio. & xresnet1d50& 0.8295(27) & 0.9260(25) \\
\bottomrule
    \end{tabular}

\end{table*}

We start by discussing the linear evaluation performance in \Cref{tab:summary}, which should be set in perspective to the supervised performance achieved on \textit{PTB-XL}. The best published result for this task on the same dataset with identical splits using purely supervised training was 0.925(07), also using a \textit{xresnet1d}-model \cite{Strodthoff:2020Deep}. Our supervised results remains compatible within error bars with this baseline result. The architecture used for CPC pretraining (denoted by \textit{4FC+2LSTM+2FC}) was not investigated in previous studies \cite{Strodthoff:2020Deep} and shows the strongest supervised performance reported on \textit{PTB-XL} thus far.

The linear evaluation performances in \Cref{tab:summary} show that the pretrained representations are highly relevant for downstream classification tasks. Most notably, the linear evaluation performance of the \textit{CPC} model only shows a performance gap of 0.5\% compared to the same model architecture trained in a supervised manner and already exceeds the best result previously obtained using supervised training on the same dataset \cite{Strodthoff:2020Deep}. The contrastive methods from computer vision show a slightly weaker performance, but still the best linear evaluation performance reaches 95.5\% of the respective supervised performance. The main point we are trying to convey here is how far one can push the linear evaluation performance not only in relative comparison to supervised performance, but also on an absolute scale. Whereas the former can also be demonstrated with simpler model architectures, the latter requires a certain model complexity. Based on these results, it is justified to claim that self-supervised representation learning is very effective in the ECG domain. To demonstrate the impact of the dataset size, we also report results for pretraining \textit{CPC} just on \textit{PTB-XL} i.e.\ using only 40\% of the original training dataset. As expected, increasing the size of the training dataset leads to improvements in the linear evaluation performance. Again, the reader is referred to \Cref{sec:cpc_ablations} in the supplementary material for details on the impact on the modifications in the original \textit{CPC} architecture and finetuning schedule.

\subsection{Self-supervised pretraining improves downstream performance}
\label{sec:finetuning}

In this section, we investigate whether finetuning from self-supervised representations can potentially lead to improvements in downstream performance as compared to purely supervised training. The results are compiled in \Cref{tab:summary} and summarized graphically in \Cref{fig:summary}.
As before, \textit{SimCLR} reaches the best linear evaluation performance whereas it is slightly outperformed by \textit{BYOL} in terms of downstream performance. The considerably better linear evaluation performance of the \textit{CPC} model as compared to \textit{BYOL} and \textit{SimCLR} directly translates into an improved downstream performance. Interestingly, the \textit{SimCLR} and \textit{BYOL} finetuned model performance when using physiological noise during training almost reaches the results from using (RRC,TO)-transformations, while a sizable performance gap exists between them in terms of linear evaluation performance. 
In order to test in how far the overlap between using \textit{PTB-XL} for pretraining and for evaluation leads to an overestimation of the positive effects of pretraining on unseen data, we also investigate the performance of a model pretrained on \textit{CinC2020 w/o PTB-XL}, i.e.\ \textit{CinC2020} with records from \textit{PTB-XL} excluded, which is comparable in size to \textit{PTB-XL}. On the one hand, the linear evaluation performance after pretraining on \textit{CinC2020 w/o PTB-XL} is lower and does not overlap with the result from pretraining on \textit{PTB-XL}, which supports the initial hypothesis. On the other hand, the seemingly superior representation from pretraining on \textit{PTB-XL} do not translate into a stronger classification performance after finetuning. In fact, finetuning from pretraining on \textit{CinC2020 w/o PTB-XL} even leads to a higher point estimate whereas both results remain consistent within error bars. Hence, these somewhat inconclusive results do not allow any decisive statements about the initial hypothesis. In particular, it remains difficult to disentangle the effects of overlap of samples in pretraining and finetuning from differences in the distribution of the pretraining datasets, which are rather pronounced as shown in \Cref{fig:label_distribution}.

In all cases, the results from finetuning pretrained models improve over the corresponding supervised results (by 1.0\% for \textit{CPC}, by 0.2\% for \textit{SimCLR}, and by 0.5\% for \textit{BYOL}). Furthermore, it is noticeable that already after the first finetuning step, where just the batch norm statistics and the classification head were adjusted, the \textit{CPC} model reaches performance values around 0.931, i.e.\ already roughly matches supervised performance. These results provide a clear case for self-supervised learning in the ECG domain.

At this point we find it appropriate to comment on different sources of uncertainty impacting our results. In \Cref{tab:summary}, we report uncertainties related to the inherent randomness of the training process, which we assess via multiple training runs. In addition to this systematic error, there is also an uncertainty in the final scores due to the finiteness and the particular sample distribution of the test set. As in \cite{Strodthoff:2020Deep}, we assess this error via empirical bootstrapping on the test set using 1000 bootstrap iterations and evaluate 95\% confidence intervals. For finetuning after self-supervised pretraining we find confidence intervals $\pm0.006$ ($\pm0.008$ for linear evaluation) as opposed to $\pm0.008$ for training from scratch with only minor variations between different runs. This statistical error therefore represents the dominant source of uncertainty, which provides a strong argument for larger ECG evaluation datasets.
As in \cite{Strodthoff:2020Deep}, we check if the confidence interval for the difference between finetuning following pretraining and training from scratch encloses only positive values, which would indicate that the performance improvement is statistically significant. We investigated the improvements for every combination of the 10 models that were finetuned from the model pretrained on \textit{All} and 10 models trained in a supervised fashion, combining both sources of uncertainty in a single analysis. In 90\% of these 100 comparisons, pretraining led to a statistically significant improvement, underscoring the positive effects of self-supervised pretraining.

\begin{figure}[t]
	\centering
	\includegraphics[width=\columnwidth]{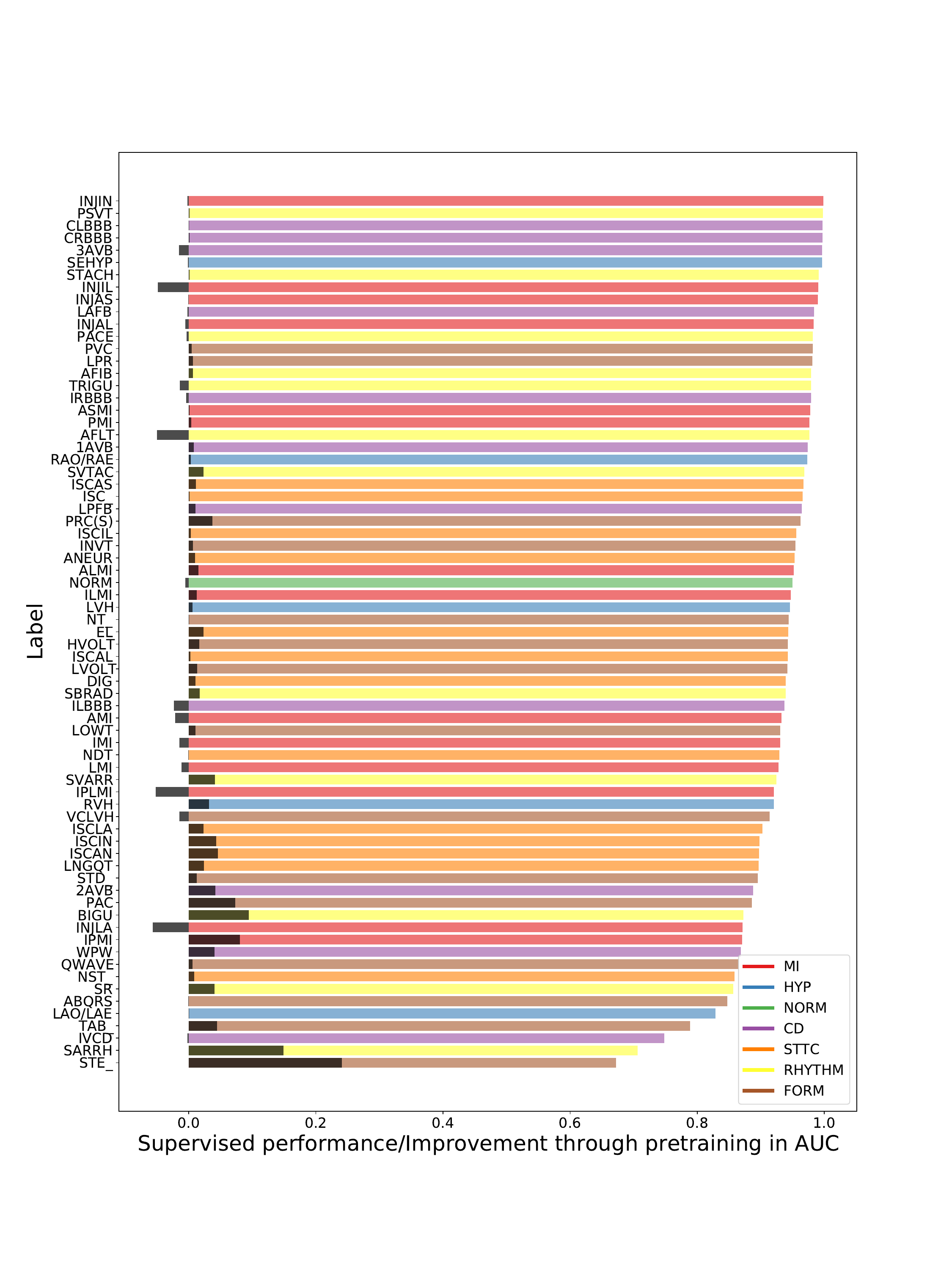}
	\caption{Individual label AUCs for a \textit{4FC+2LSTM+2FC}-model from supervised training (in color, sorted in descending order) and the corresponding improvements through self-supervised training with \textit{CPC} (in black). }
	\label{fig:sup_performance}
\end{figure}

In \Cref{fig:sup_performance}, we break down the macro AUC values of the best performing model by ECG statement. Per statement, we present both the performance of the supervised trained model (in color) and the improvement (in black) that occurs through self-supervised training. Moreover, we color the statements by their respective super class (for diagnostic statements), or mark them as rhythm or form statement and sort them according to the supervised performance. In terms of improvements through self-supervised pretraining, the largest average gains are observed for form and rhythm statements, see \Cref{fig:super_improv}, which supports the hypothesis that different pathologies tend to profit differently from self-supervised pretraining. This effect is superimposed by a different effect that is visible from \Cref{fig:sup_performance}, which indicates that the gains through pretraining show a negative correlation with the performance level before pretraining, i.e.\ ECG statements with low supervised performance tend to profit more from self-supervised pretraining, see \Cref{fig:improv} for an explicit demonstration.

As a final remark, one has to consider the different sizes of models under consideration. Whereas the \textit{CPC}-model during finetuning comprises 5.8M parameters, the \textit{xresnet1d50} only counts 930K parameters, which might suggest that part of the gap between \textit{CPC} and \textit{BYOL/SimCLR} can at least partially be attributed to a difference in model capacity. However, in preliminary experiments we saw no indications of strong performance increases with wider or deeper models. It remains to see how the performance of \textit{SimCLR} and \textit{BYOL} scales on larger datasets. On the contrary, for the \textit{4FC+2LSTM+2FC}-models, the model capacity is instrumental to reach its very high performance, see \Cref{sec:cpc_ablations}. A serious disadvantage of \textit{CPC} is the sequential nature of the LSTM, which leads to slow training times. The training times are not directly comparable due to the different nature of the tasks, but give at least a hint. \textit{CPC} models were pretrained for 200 epochs, which took approximately 6 days on a single Tesla V100 GPU. \textit{SimCLR} and \textit{BYOL} pretraining was performed for 2000 epochs using batch sizes of 8192 with approximate runtimes of 15h and 13h on a single Tesla V100 GPU, respectively. Performing 50+20(100) epochs of finetuning for the \textit{4FC+2LSTM+2FC}(\textit{xresnet1d50})-model takes approximately 25(10)~minutes on the mentioned hardware.

\begin{figure}[t]
	\includegraphics[width=\columnwidth]{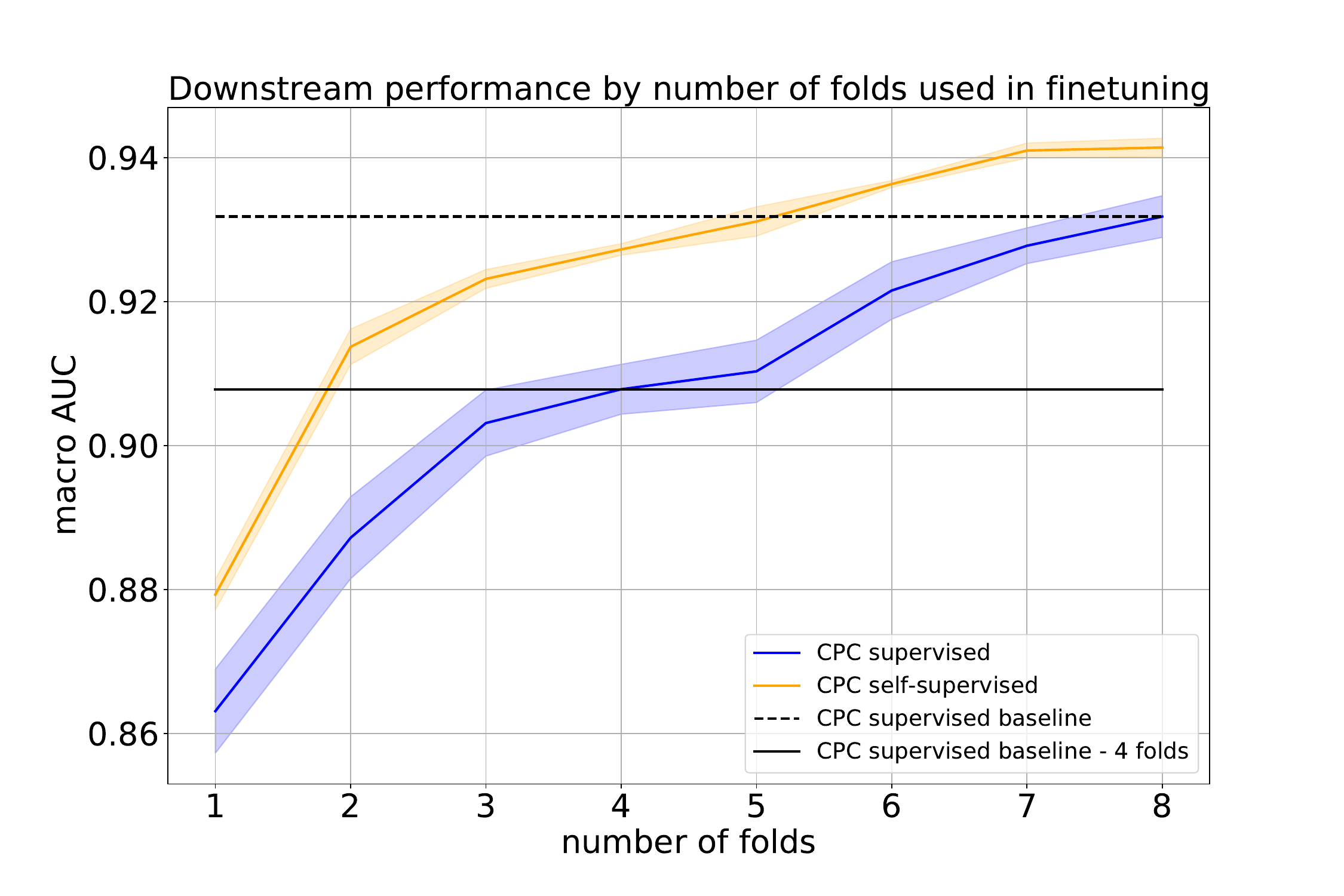}
	\caption{Finetuning downstream performance on PTB-XL dataset of a \textit{4FC+2LSTM+2FC}-model pretrained using \textit{CPC} compared to its supervised counterpart. The finetuning was performed for different number of training folds, ranging from 1-8 folds. We used 10 runs for 8 folds as before and 3 runs for 7 folds or fewer. The plot shows the mean performance as a solid line and one standard deviation around it as a shaded band. To guide the eye, we indicate the performance of the supervised model trained on 8(4) folds performance on the full training set} by a dashed(solid) line .
	\label{fig:trained_on_folds}
\end{figure}

\subsection{Self-supervised pretraining improves downstream data efficiency}
\label{sec:dataefficiency}
Another potential advantage of self-supervised pretraining lies in a potentially improved data efficiency when finetuned on a downstream task. This is a particularly relevant case for medical applications, where high-quality labels are hard to obtain. To investigate this claim in detail, we compare the performance of different pretrained models from self-supervised representations to models trained from scratch in purely supervised fashion while varying the number of training folds from the original 8 folds to a single fold. This can be read off for example from the number of training folds where the pretrained model reaches the same performance as the supervised model trained on the full dataset. In the case of \textit{CPC}, this point is reached approximately at 5 folds or equivalently approximately 62\% of the training data. The performance level of the supervised models at 4 folds is approximately reached by the pretrained model trained only on two folds i.e.\ 50\% of the training data. For \textit{BYOL} and \textit{SimCLR}, the effect is still present but much less pronounced due to the closer proximity of the pretrained and the purely supervised results on the full training set. \Cref{fig:trained_on_folds} nicely illustrates another advantage of self-supervised pretraining, namely the performance across different training runs is much more stable compared to purely supervised training, as visible from considerably reduced error bands, most crucially influenced by the two-step finetuning procedure in combination with discriminative learning rates, see also \Cref{sec:cpc_ablations} in the supplementary material.

\begin{figure}[t]
	\includegraphics[width=\columnwidth]{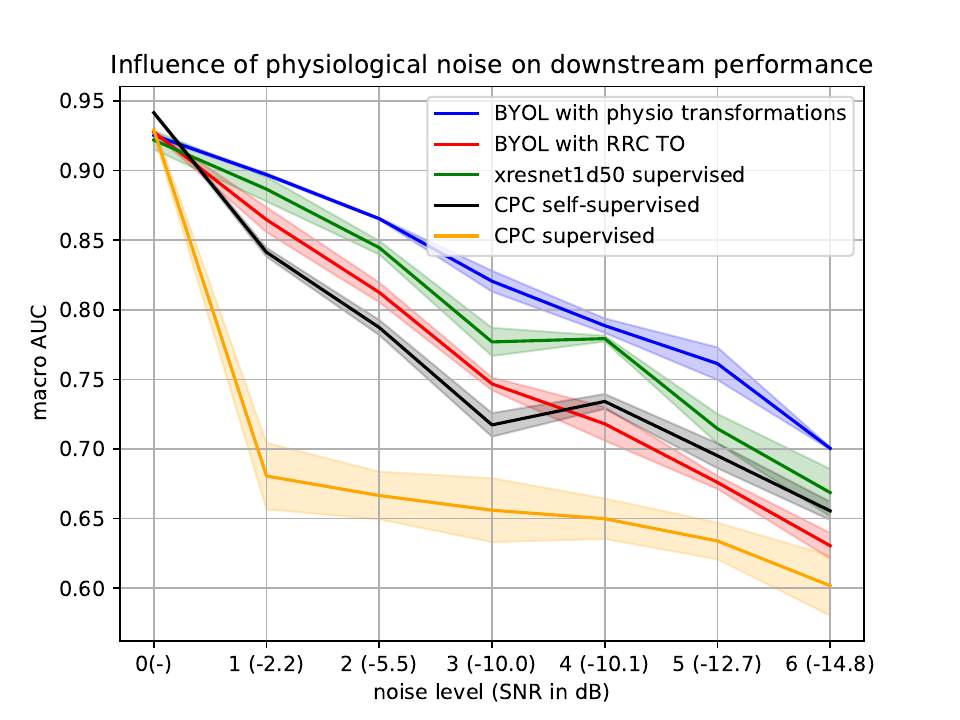}
	\caption{Evaluating the impact of noise on pretrained and purely supervised trained classifiers. We induced typical physiological noise at different strength levels to the test set.}
	\label{fig:noise}
\end{figure}

\subsection{Self-supervised pretraining improves robustness of downstream classifiers}
\label{sec:robustness}
In addition to quantitative performance and data efficiency, robustness is one of the key quality quality criteria for machine learning models. Here, we focus on robustness against input perturbations. It is well known that certain types of noise tend to occur in ECG data as a consequence of the measurement process and physiological interference \cite{Friesen1990,Lenis2017}. In \Cref{sec:transformations} in the supplementary material, we briefly review typical kinds of ECG noise and propose simple ways of parameterizing them. For simplicity, we just superimpose the different noise types and the original ECG waveform. We define different noise levels by adjusting the amplitudes of these noise transformations and evaluate the performance of the models from the previous sections on perturbed versions of the original test set. We also indicate signal-to-noise-ratios (SNRs) corresponding to the different noise levels, where we identify the signal with the original ECG waveform. However, one has to keep in mind that this assessment neglects the noise inherent in the original measurement, which implies that the given SNRs only upper-bound the actual values.

The goal is to investigate if pretrained models are less susceptible to physiological noise. The results in \Cref{fig:noise} reveal an interesting pattern: For the \textit{4FC+2LSTM+2FC}-models, the \textit{CPC}-pretrained model shows a considerably improved robustness to noise compared to its supervised counterpart. However, both models turn out to be less robust than the considerably less complex \textit{xresnet1d}-models. For the latter, the \textit{BYOL}-pretrained model with physiological noise shows the strongest overall performance and also performs considerably stronger than the corresponding supervised model, which is to a certain degree an expected result as the model experienced a comparable kind of noise during pretraining. This result provides a strong argument for pretraining with domain-specific noise transformations even if it comes at the cost of a slight performance loss compared to the best-performing pretraining procedure during noiseless evaluation, see \Cref{tab:summary}. Somewhat surprisingly, the \textit{BYOL}-pretraining with the artificial (RRC,TO)-transformation even lead to a reduced robustness compared to the supervised \textit{xresnet1d50}. As a final remark, the noise levels 3 and beyond are already strongly dominated by noise and correspond to situations that will rarely be encountered in real-world scenarios. 

\subsection{Implications of the results}
Even though we discussed the implications of our findings on a technical level, we find it appropriate to also briefly comment on the broader implications of our results. \Cref{sec:lineareval} addresses the question how far one can push the linear evaluation performance and demonstrates that self-supervised learning can be implemented effectively in the domain of ECG data. In terms of (clinical) impact, however, it is primarily a necessary prerequisite for the following investigations: \Cref{sec:finetuning} shows that self-supervised pretraining improves the predictive performance compared to training from scratch. Most importantly, breaking down the improvements according to individual diagnoses shows that underperforming diagnoses when training from scratch profit most from self-supervised pretraining. This is an encouraging sign as it entails the prospects to eventually train models on even finer and hence less populated label hierarchies and/or to tackle rare diseases, for which only a very limited number of labeled samples exist in the first place. \Cref{sec:dataefficiency} demonstrates that the improvements achieved through pretraining directly translate into an improved label efficiency. This is also an encouraging result for the broader ECG research community given the growing but still compared to commercial ECG datasets small sample size available from freely accessible ECG datasets with high label quality. Finally, \Cref{sec:robustness} stresses that self-supervised pretraining leads benefits such as improved robustness that go beyond quantitative performance and that are also very desirable in clinical applications.
	
	\section{Summary and conclusions} 
In this work, we put forward a comprehensive assessment of self-supervised representation learning on 12-lead clinical ECG data. Even though self-supervised algorithms have been applied successfully in computer vision and speech, ECG is a different data modality, where the degree to which self-supervised learning works is not clear from the onset. Upon adapting self-supervised representation learning to the ECG domain, self-supervised learning turns out to be very effective: Self-supervised representations (via \textit{CPC}) reach scores that only fall behind 0.5\% supervised performance during linear evaluation and lead to improvements of  1.0\% compared to supervised performance during finetuning, which represents a significant increase in 90\% of the time within an analysis incorporating both systematic as well as statistical uncertainties. The sizable performance gap translates into an improved label efficiency, i.e.\ the pretrained model reaches the same performance as the supervised model but using only roughly 50-60\% of the samples.
We also investigate the impact of self-supervised pretraining on the robustness of the corresponding finetuned classifiers against physiological noise. We find increased robustness for most pretrained models compared to the corresponding models trained from scratch, but particularly for those that were pretrained using domain-specific noise transformations. This provides a strong case for the use of domain-specific noise transformations during pretraining. 

To summarize, self-supervised learning is one path towards more robust and more label-efficient training procedures, which might alleviate the problem of label scarcity, which is particularly pressing in medical applications. In this work, we demonstrated clear advantages in terms of quantitative performance, label efficiency and robustness. It will be interesting if these carry over to further quality dimensions. We see our work as a first step towards measurable progress in the field of representation learning for 12-lead ECGs. All used datasets and the source code underlying our study are publicly available \cite{implementation}.

\section*{Declaration of competing interest}
The authors declare that they have no known competing financial interests or personal relationships that could have appeared to influence the work reported in this paper.

\section*{Acknowledgments}
This work was supported by the Bundesministerium f\"ur Bildung und Forschung through the BIFOLD - Berlin Institute for the Foundations of Learning and Data (ref. 01IS18025A and ref. 01IS18037A) and 18HLT07 MedalCare. The project 18HLT07 MedalCare has received funding from the EMPIR programme co-financed by the participating states and from the European Union's Horizon 2020 research and innovation programme.

	\bibliography{bibfile}    
    \bibliographystyle{IEEEtran}
    
\renewcommand{\thesubsection}{\Alph{subsection}.\arabic{subsection}}
\setcounter{subsection}{0}
    		\clearpage
       
        \begin{appendices}
        	\renewcommand{\thesubsection}{\fnsymbol{subsection}}
        	    	
\section{Transformations used during pretraining}
\label{sec:transformations}
In this section, we address the transformations used for the presented computer vision-based self-supervised learning methods (SimCLR, BYOL, SwAV). We distinguish between artificial transformations (\Cref{sec:a1}), which are partially adapted versions of the transformations used in \cite{chen2020simple} or transformations specific to time series, and physiological transformations (\Cref{sec:a2}), which are more realistic perturbations that occur due to inaccuracies in the measurement process.\\
\Cref{fig:artificial} and \Cref{fig:physio} depict single-lead examples for each of the artificial and physiological transformations, respectively.

\subsection{Artificial transformations}
\renewcommand{\thefigure}{A\arabic{figure}}
\setcounter{figure}{0}
\label{sec:a1}
\paragraph{Gaussian noise} \textit{Gaussian noise} describes the addition of zero-mean Gaussian noise to all channels. The standard deviation $\sigma$ of the noise is the only parameter of the transformation. We used $\sigma=0.01$~mV in our experiments.
\paragraph{Gaussian blur} \textit{Gaussian blur} describes the application of a one-dimensional Gaussian kernel $k$ to the ECG signal, which results in a blurred version of the signal. More specifically, We used a kernel with entries $(0.1, 0.2, 0.4, 0.2, 0.1)$ in our experiments.
\paragraph{Channel resize} \textit{Channel resize} multiplies the $i$-th channel of the signal by the factor $c_i=b^a_i$, where $b$ is the only parameter of the transformation and $a_i$ is uniformly sampled from $[-1, 1]$, such that $\mathbb{E}[c_i] = 1$. In our experiments we chose $b=3$. \textit{Channel resize} can be seen as an analogue of color transformations in computer vision.
\paragraph{Random resized crop} \textit{Random resized crop} crops a random contiguous segment of the signal and rescales it to its original size. We sample the crop parameter $p$ uniformly from the range $(l, m)$, where $(l, m)$ are the parameters of the transformation. In our experiments we used $(l, m)=(0.5, 1.0)$, that is we cropped the signals to portions between 50\%-100\%. 
\paragraph{Time out} \textit{Timeout} \cite{cheng2020subjectaware} sets a random contiguous segment of the signal to zero. It accepts as parameters a range $(t_l, t_u)$, from which the timeout parameter $t$ is uniformly sampled. The parameter describes how much of the signal will be set to zero. In our experiments, we used $(t_l, t_u)=[0.0, 0.5]$, therefore we set up to 50\% of the signal to zero.
\paragraph{Dynamic time warp} \textit{Dynamic time warp} stretches and squeezes random contiguous segments of the signal along the x-axis. The parameters are the number of warps $w$ and the radius $r$ of the warps (in timesteps). We used $w=3$ and $r=10$.

\begin{figure*}[t!]
	\centering
	\begin{subfigure}[b]{0.4\textwidth}
		\centering
		\includegraphics[width=\textwidth]{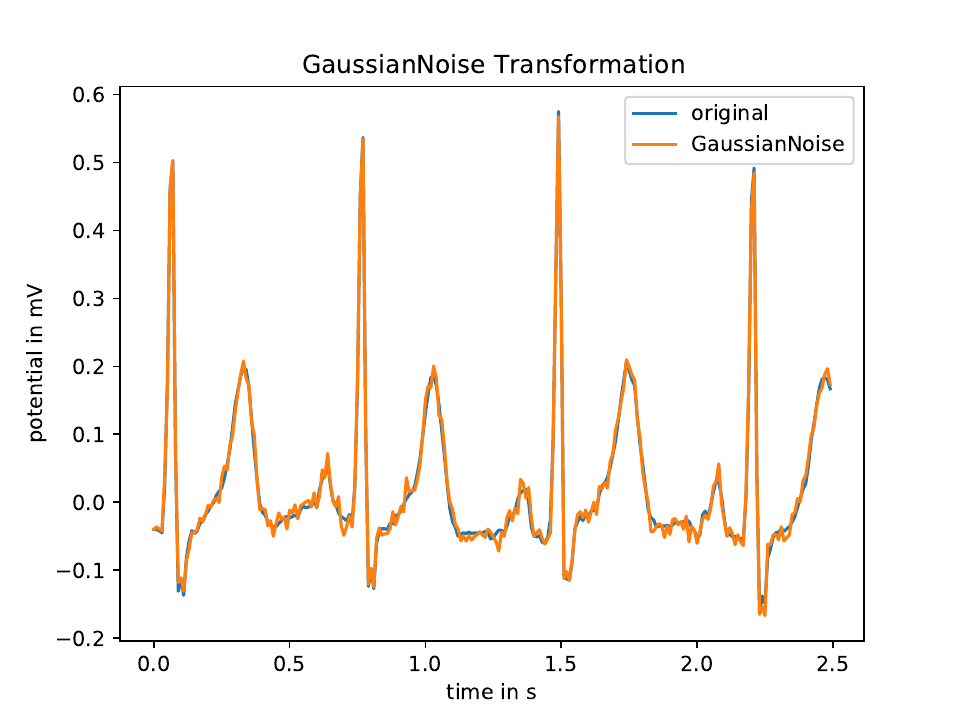}
		\caption{}
		\label{fig:a}
	\end{subfigure}
	\begin{subfigure}[b]{0.4\textwidth}
		\centering
		\includegraphics[width=\textwidth]{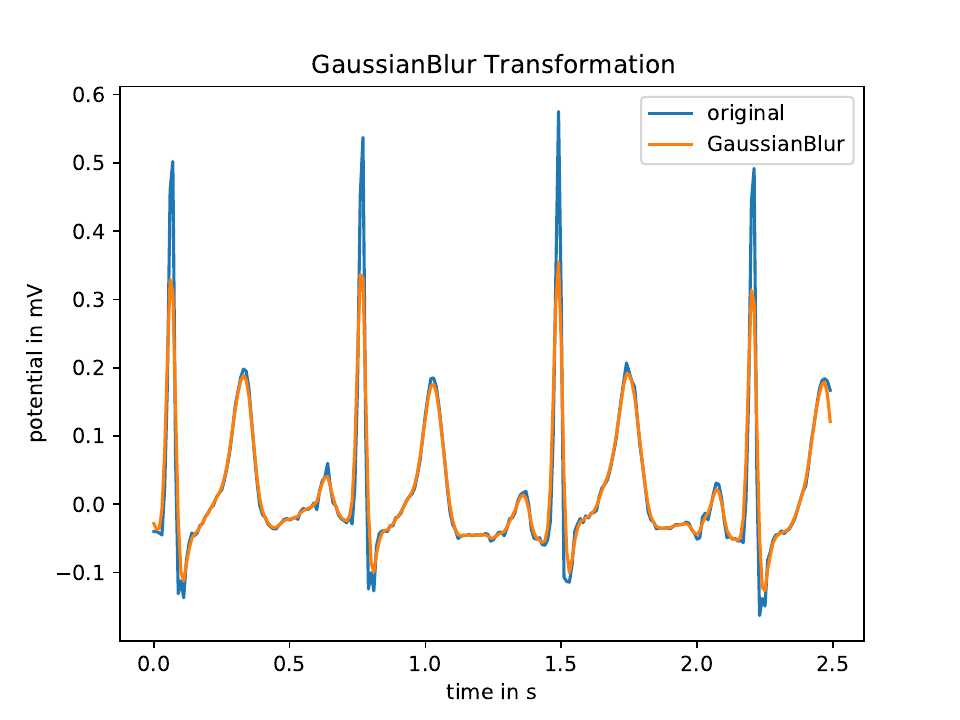}
		\caption{}
		\label{fig:b}
	\end{subfigure}
	\begin{subfigure}[b]{0.4\textwidth}
	\centering
	\includegraphics[width=\textwidth]{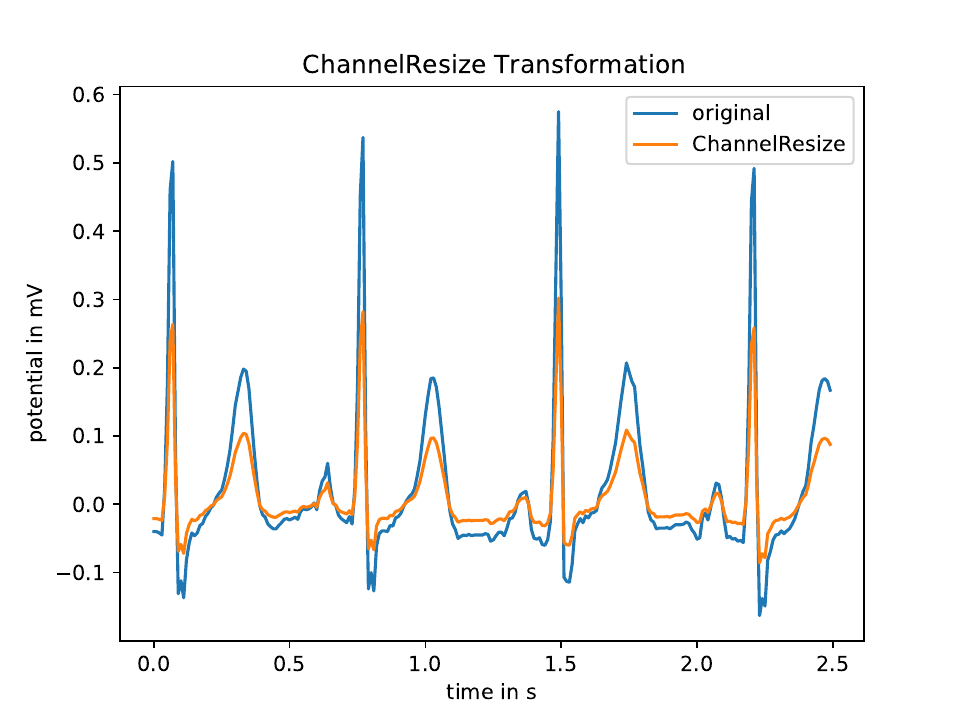}
	\caption{}
	\label{fig:c}
\end{subfigure}
\begin{subfigure}[b]{0.4\textwidth}
	\centering
	\includegraphics[width=\textwidth]{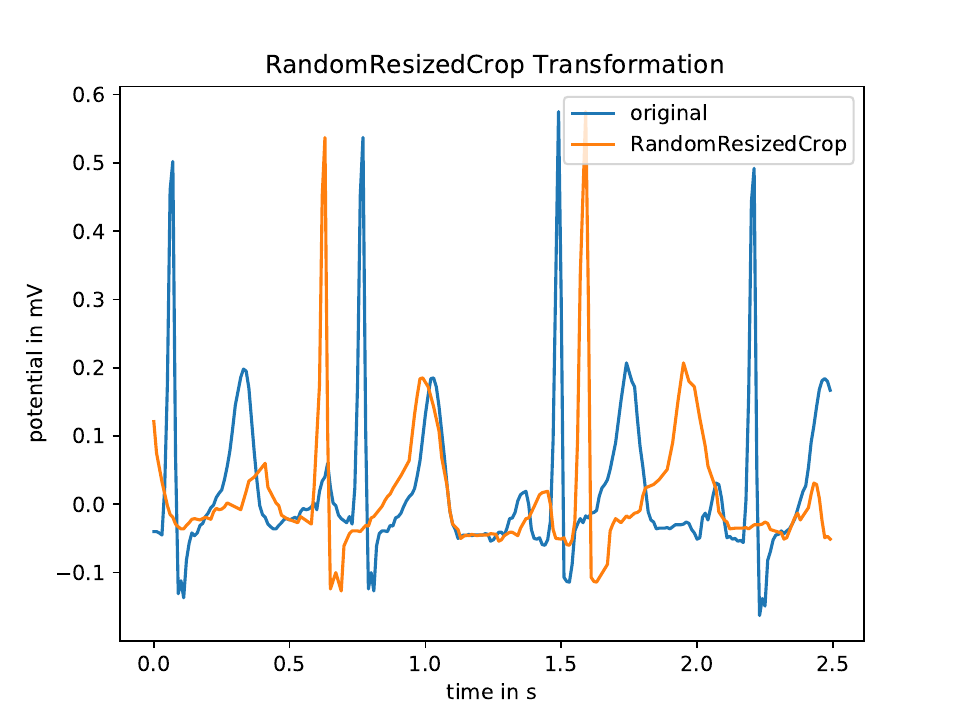}
	\caption{}
	\label{fig:d}
\end{subfigure}
	\begin{subfigure}[b]{0.4\textwidth}
	\centering
	\includegraphics[width=\textwidth]{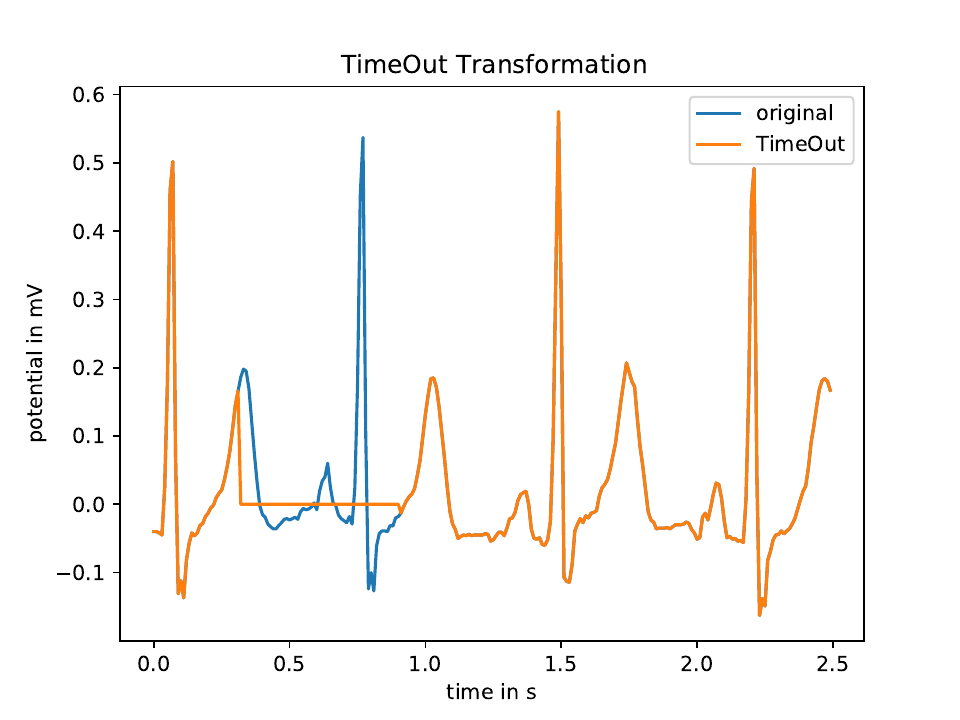}
	\caption{}
	\label{fig:e}
\end{subfigure}
\begin{subfigure}[b]{0.4\textwidth}
	\centering
	\includegraphics[width=\textwidth]{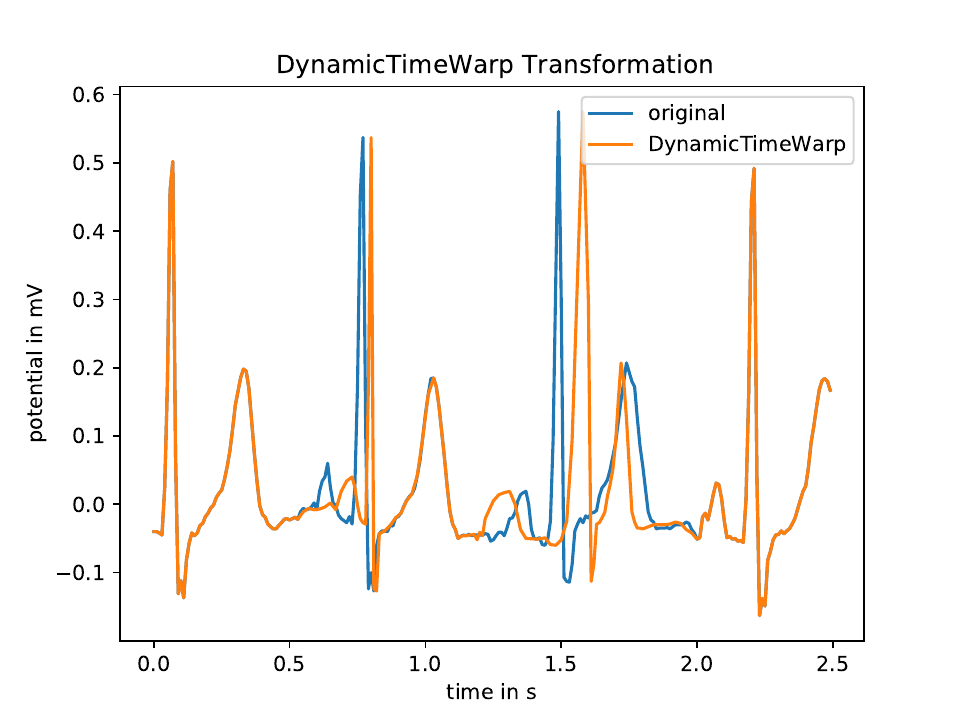}
	\caption{}
	\label{fig:f}
\end{subfigure}
	\caption{Artificial Transformations used for computer vision based self-supervised methods: (a) \textit{Gaussian noise}, (b) \textit{Gaussian blur}, (c) \textit{Channel resize}, (d) \textit{Random resized crop} and (e) \textit{Time out}.}
	\label{fig:artificial}
\end{figure*}

\subsection{ECG-specific physiological noise transformations}
\renewcommand{\thefigure}{A\arabic{figure}}
\setcounter{figure}{0}
\label{sec:a2}
\begin{figure*}[t!]
	\centering
	\begin{subfigure}[b]{0.4\textwidth}
		\centering
		\includegraphics[width=\textwidth]{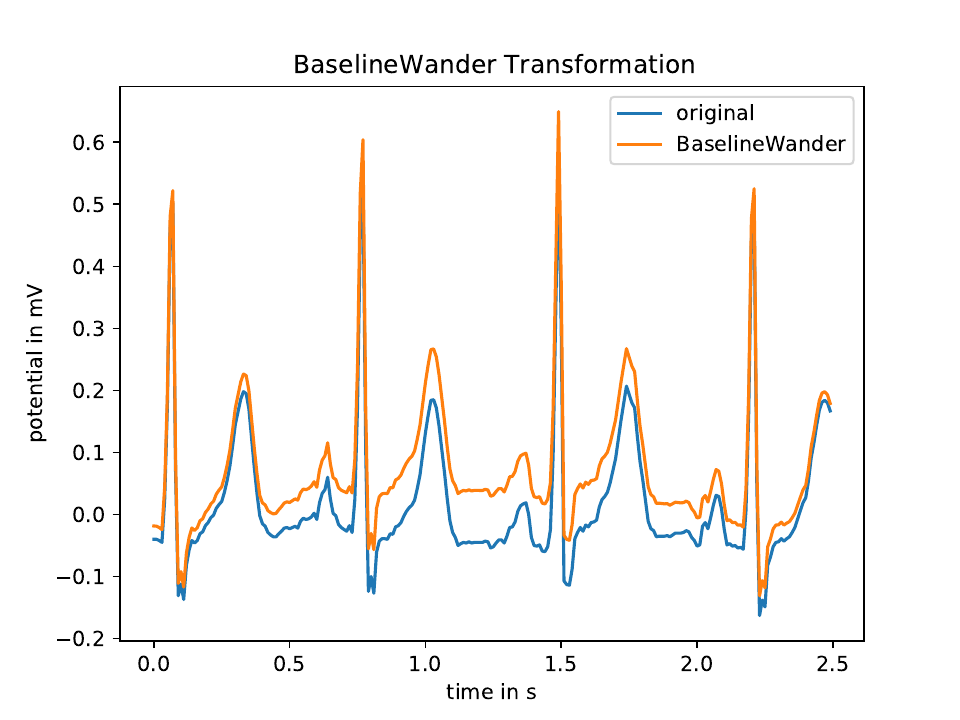}
		\caption{}
		\label{fig:a2}
	\end{subfigure}
	\begin{subfigure}[b]{0.4\textwidth}
		\centering
		\includegraphics[width=\textwidth]{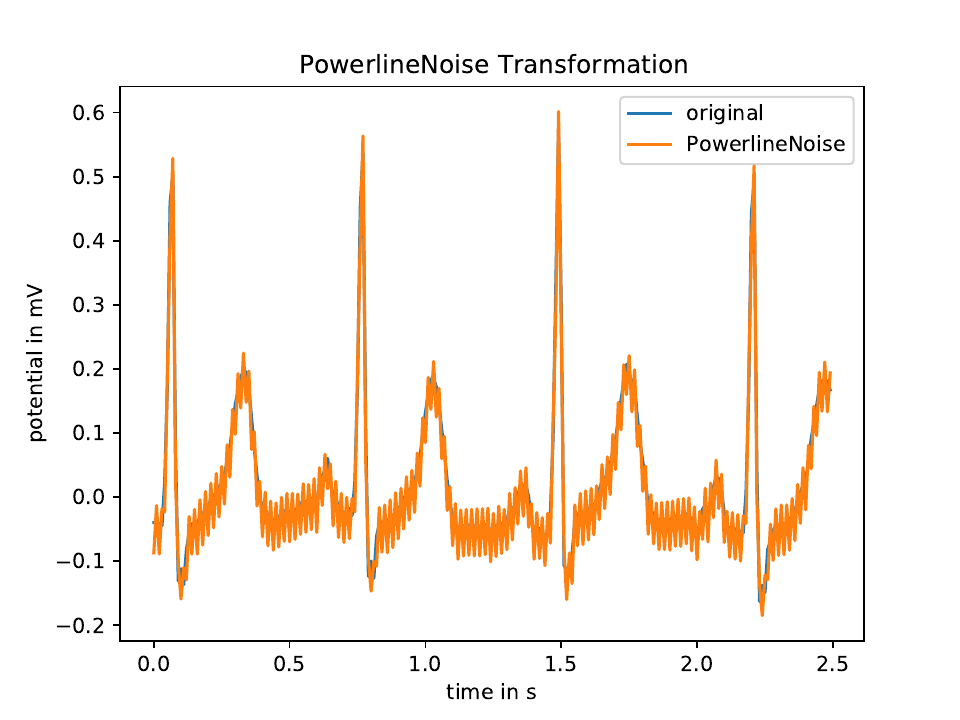}
		\caption{}
		\label{fig:b2}
	\end{subfigure}
	\begin{subfigure}[b]{0.4\textwidth}
		\centering
		\includegraphics[width=\textwidth]{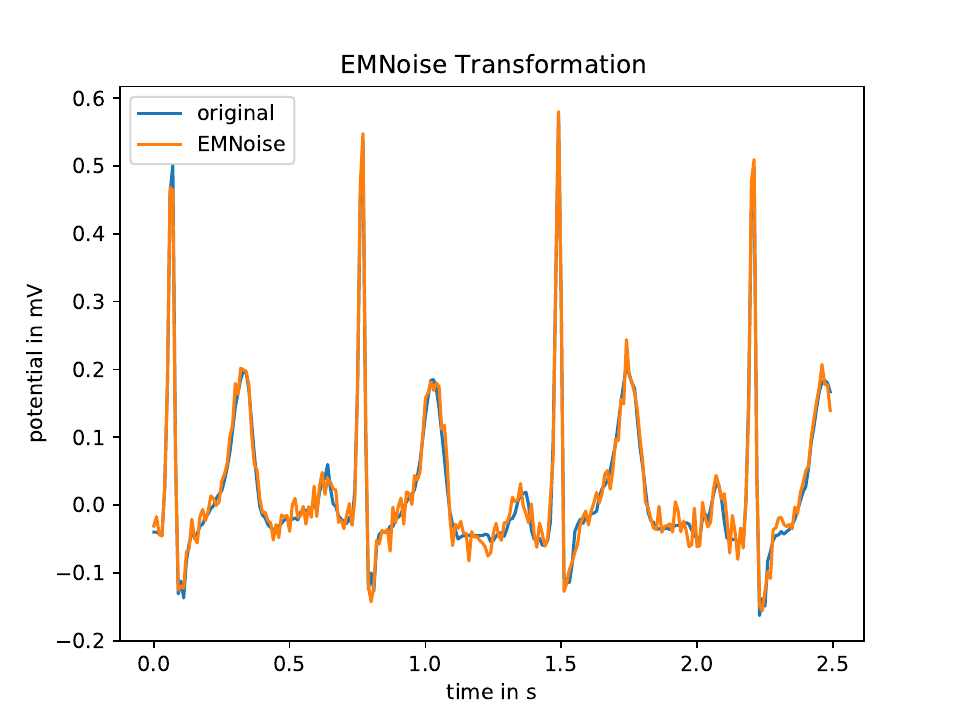}
		\caption{}
		\label{fig:c2}
	\end{subfigure}
	\begin{subfigure}[b]{0.4\textwidth}
		\centering
		\includegraphics[width=\textwidth]{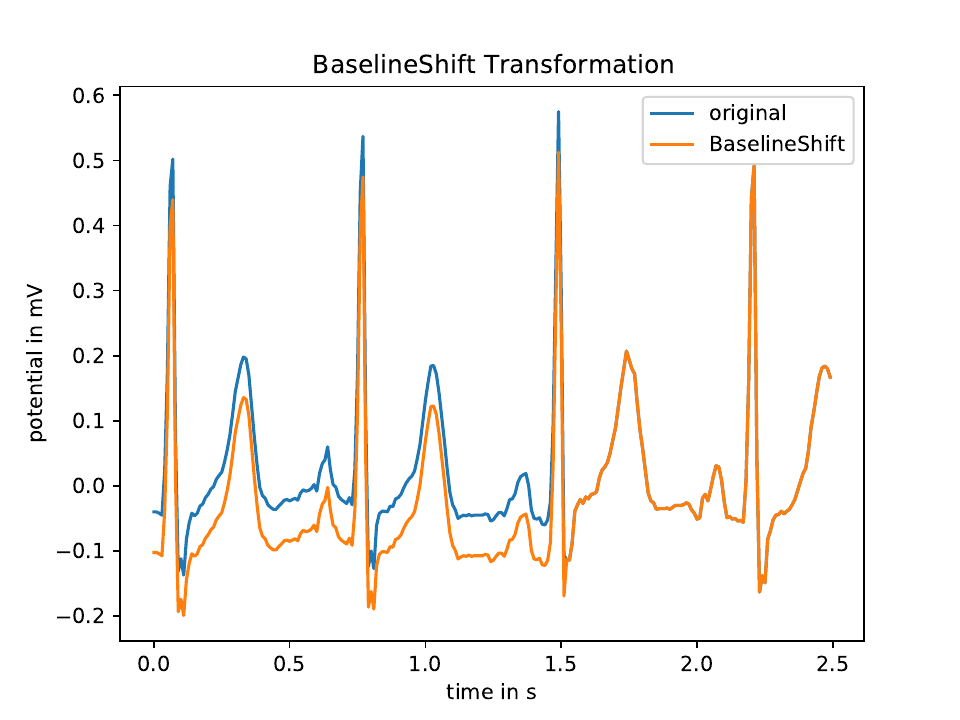}
		\caption{}
		\label{fig:d2}
	\end{subfigure}
	\caption{ECG-specific physiological noise transformations used in computer vision based self-supervised methods: (a) \textit{Baseline wander}, (b) \textit{Powerline noise}, (c) \textit{Electromyographic noise} and (d) \textit{Baseline shift}.}
	\label{fig:physio}
\end{figure*}

\paragraph{Baseline wander} Baseline wander is a low-frequency ECG artifact that arises from respiration, electrically charged electrodes or movement of the patient. Here, we follow \cite{Lenis2017} and model it as a superposition of different sinusoidal components:
\begin{equation}
    n_\text{blw}(t)_i= C c_i \sum_{k=1}^K a_k \cos(2\pi t k \Delta f +\phi_k)\,,
\end{equation}
where $C$,$a_k$,$\phi_k$ are uniform random numbers with ranges $[0,C_\text{max,blw}]$, $[0,1]$, and $[0,2\pi]$. We use $\Delta f=0.01$~Hz, $f_c=0.05$~Hz and $K= \left \lfloor{\Delta f/f_c}\right  \rceil$. $c_i$, where the index $i$ designates the $i$th lead, is drawn from a standard normal distribution and modulated by a random sign.

\paragraph{Powerline noise} Powerline noise describes powerline interference pickup at $f_n=50$~Hz and its higher harmonics \cite{Friesen1990}. Here, it is modeled as 
\begin{equation}
    n_\text{pln}(t)_i= C c_i \sum_{k=1}^K a_k \cos(2\pi t k f_n +\phi_1)\,,
\end{equation}
with $K=3$ and variables as defined above except for $c_i$, which is drawn from a uniform distribution over $[-1,1]$ in this case and $C$, which is a uniform random number drawn from $[0,C_\text{max,pln}]$.

\begin{table}[]
	\centering
	\caption{Mapping of noise levels to corresponding parameters of physiological transformations.}
	
	\begin{tabular}{l|lllll}
		\toprule
		noise level & $C_\text{max,blw}$ & $C_\text{max,pln}$& $C_\text{max,emn}$& $C_\text{max,bls}$& SNR\\
		
		\midrule
		1 & 0.05 & 0.25 & 0.1 & 0.5 & -2.2 \\\midrule
		2 & 0.1 & 0.5 & 0.2 & 1 & ~-5.5\\\midrule
		3    & 0.1 & 1 & 0.2 & 2 & ~-10.0 \\\midrule
		4    & 0.2 & 1 & 0.4 & 2 & ~-10.1\\\midrule
		5   & 0.2 & 1.5 & 0.4 & 2.5 & ~-12.7\\\midrule
		6    & 0.3 & 2 & 0.5 & 3 & ~-14.8\\\midrule
		\bottomrule
	\end{tabular}
	\label{tab:noise}
\end{table}
\paragraph{Electromyographic noise} Electromyographic describes high-frequency noise typically caused by muscle contractions \cite{Friesen1990}. Here, we simply model it as Gaussian noise:
\begin{equation}
    n_\text{emn}(t)_i= \beta\,,
\end{equation}
where $\beta$ is drawn from a normal distribution with mean zero and variance $C_\text{max,emn}$.

\paragraph{Baseline shift} Baseline shift describes baseline changes through electrode-skin impedance changes through electrode motion \cite{Friesen1990}. Following \cite{Friesen1990}, we model this type of noise by sampling a stepwise function $\text{swf}(t)$. In our case, it is created as follows: We determine the number of segments by drawing a random integer from $[0,\left \lceil{\text{bls}_\text{max}\cdot L/f_s}\right  \rceil]$, where $\text{bls}_\text{max}=0.3 s^{-1}$, $L$ is the length of the segment (in timesteps) and $f_s$ is the sampling frequency. For each segment, we add a step function with non-zero values at a segment with length drawn from a normal distribution with mean $f_s \text{bls}_\text{len,mean}$ and standard deviation $0.2\cdot f_s\cdot \text{bls}_\text{len,mean}$, where  $\text{bls}_\text{len,mean}=3$s. The amplitude of each segment determined in this way is drawn from a random uniform distribution. Given, the sampled stepwise constant function $\text{swf}(t)$, one defines the baseline shift noise via
\begin{equation}
    n_\text{bls}(t)_i= C c_i \text{swf}(t)\,,
\end{equation}
where $C$ is a uniform random number drawn from $[0,C_\text{max,bls}]$ and  $c_i$ is drawn from a standard normal distribution and modulated by a random sign.

\paragraph{Superposition} Eventually, all four noise types are superimposed and added to the original signal $s(t)$ i.e.\
\begin{equation}
    s_\text{physio. noise}(t) = s(t) + n_\text{blw}(t) + n_\text{pln}(t) + n_\text{emn}(t) + n_\text{bls}(t)
\end{equation}
The noise strength is adjusted by varying $C_\text{max,blw}$, $C_\text{max,pln}$, $C_\text{max,emn}$, and $C_\text{max,bls}$ while keeping all other parameters fixed.

\subsection{Parameter values used during pretraining and evaluation}

During pretraining, we used $C_\text{max,blw}=0.1$, $C_\text{max,pln}=0.2$, $C_\text{max,emn}=0.5$, and $C_\text{max,bls}=1$ when using the physiological transformations. For our robustness test, we created noisy validation sets. We considered different levels of noise, which are described in Table \ref{tab:noise}.

\section{SimCLR, BYOL, and SwAV and augmentation transformations}
\label{sec:model_selection}
\renewcommand{\thefigure}{B\arabic{figure}}
\setcounter{figure}{0}

\begin{figure}[t]
	\includegraphics[width=\columnwidth]{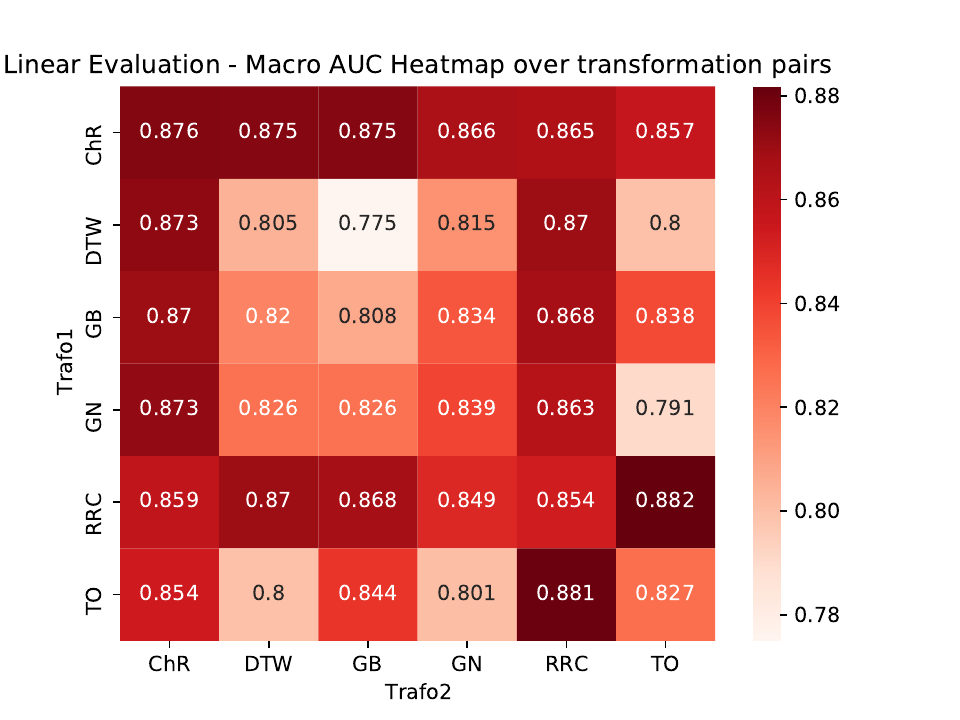}
	\caption{Linear evaluation performance (macro AUC) on the \textit{PTB-XL} validation set of a \textit{xresnet1d50} model after 500 epochs pretraining on \textit{All} with \textit{SimCLR} using one or two data augmentations. Diagonal entries correspond to a single transformation and off-diagonal entries correspond to the sequential composition of two transformations. We report the mean over three linear evaluation runs. }
	\label{fig:lin_heatmap}
\end{figure}

The choice of appropriate transformations to induce two semantically equivalent views on the original instance is crucial for the effectiveness of the approach and one of the key components for the success of \textit{SimCLR} \cite{chen2020simple}. In order to find the best combination of transformations during pretraining, we followed the example of \cite{chen2020simple} and performed a grid search, based on six transformations, which are partly inspired from computer vision and partly from time series analysis: \textit{Gaussian noise (GN)}, \textit{Gaussian blur (GB)}, \textit{channel resize (CR)}, \textit{time out (TO)}, \textit{random resized crop (RRC)} and \textit{dynamic time warp (DTW)}, see \Cref{sec:transformations} for a detailed description. For all pairs as well as single transformations, we trained a \textit{xresnet1d50} using \textit{SimCLR} for 500 epochs on the \textit{All} dataset. \Cref{fig:lin_heatmap} shows the respective linear evaluation performances on the \textit{PTB-XL} dataset. The results rather clearly identify \textit{time out} in combination with \textit{random resized crop} as most effective transformation pair. This is one combination of transformation that will be used for all following experiments. For comparison, we train a model on transformations that are supposed to mimic common types of physiological noise that typically occur in ECG measurements \cite{Friesen1990,Lenis2017}. Here, we consider \textit{baseline wander}, \textit{powerline noise}, \textit{electromyographic noise} and \textit{baseline shift}, which are also described in detail in \Cref{sec:transformations}.

\begin{table}[t]
	\centering
	\caption{Comparing different contrastive learning frameworks and augmentation transformations in terms of linear evaluation and finetuning performance after 2000 epochs pretraining on the \textit{All} dataset. We report mean and standard deviation of the validation set scores over 10 finetuning runs using a concise error notation where e.g.\ 0.8976(11) signifies $0.8976\pm 0.0011$.}
	\label{tab:model_selection}
	\begin{tabular}{ll|ll}
		\toprule
		method & transformations & \multicolumn{2}{c}{PTB-XL}\\
		
		&   & lin. eval. & finetuned \\
		\midrule
		SimCLR & (RRC, TO) & \textbf{0.8976(11)} & 0.9294(14) \\
		SimCLR & physio. & 0.7957(23) & 0.9290(13)\\\midrule
		BYOL    & (RRC, TO) & 0.8781(24) & \textbf{0.9327(20)} \\
		BYOL    & physio.  & 0.8483(24)& 0.9289(20)\\\midrule
		SwAV    & (RRC, TO) & 0.8227(11) & 0.9227(19)\\
		SwAV    & physio. & 0.7157(27)& 0.9260(28)\\\midrule
		\bottomrule
	\end{tabular}

\end{table}

In a second step, we aim to identify the most effective pretraining framework. To this end, we pretrain models using \textit{SimCLR}, \textit{BYOL}, and \textit{SwAV} each with (RRC, TO) and physiological noise transformations. The results are summarized in \Cref{tab:model_selection} both in terms of linear evaluation as well as finetuned performance. As first observation, in terms of both evaluation modes the models pretrained with artificial transformations are considerably stronger than their counterparts pretrained using physiological noise. In terms of linear evaluation performance, the gap is smallest in case of \textit{BYOL}, which is consistent with findings about a less pronounced sensitivity to transformation choices in computer vision \cite{grill2020bootstrap}. However, the most interesting observation is the mismatch between linear evaluation and finetuned model scores: Whereas \textit{SimCLR} reaches clearly the best linear evaluation performance, finetuning from a \textit{BYOL} representation leads to a superior downstream performance after finetuning. This iterates the fact that the ranking in terms of linear evaluation performance is not necessarily a perfect proxy for the ranking in terms of downstream performance.

\section{CPC ablation studies and comparison to speech}
\label{sec:cpc_ablations}
In this final section, we investigate the impact of different modifications of the \textit{CPC} architecture and training procedure to demonstrate in how far they positively impacted the performance. These results potentially convey general insights for \textit{CPC} and related self-supervised approaches that go beyond the specific application to ECG data. Therefore, we vary one aspect while keeping the other ones fixed and report the impact on linear evaluation and finetuning procedure. Pretraining and evaluation is performed on \emph{PTB-XL} for simplicity. We refer to the configuration with fully connected encoder, MLP during pretraining, predicting 12 steps ahead, hidden layer, batch normalization and dropout in the classification head, two-step finetuning, discriminative learning rates during finetuning as \textit{CPC Baseline}. 

\begin{table}[hb]
	\centering
	\caption{Impact of different architectural and procedural components during \textit{CPC} pretraining and finetuning. We report the performance in comparison to our baseline result when omitting a specified component.}
	\label{tab:cpc_ablation}
	\begin{tabular}{l|ll}
		\toprule
		component & \multicolumn{2}{c}{PTB-XL}\\
		
		   & lin. eval. & finetuned \\
		\midrule
		\textit{CPC Baseline} &  \textbf{0.9226(05)} & \textbf{0.9419(13)} \\\midrule\midrule
		pretraining: no MLP & 0.9193(08) & 0.9401(13) \\\midrule
		head: no hidden layer &- & 0.9396(09) \\
		head: no BN nor dropout &- & 0.9415(11)\\
		finetuning: no two-step  &-& 0.9230(21)\\
		finetuning: no discr. lrs &- & 0.9419(21) \\
\midrule
		LSTM: 128 hidden units & 0.8799(06) & 0.9333(20)\\
		LSTM: 256 hidden units & 0.9122(04) & 0.9370(20) \\
		\bottomrule
	\end{tabular}
\end{table}

The results of this investigation are summarized in \Cref{tab:cpc_ablation}: The MLP during \textit{CPC} pretraining has a small but consistent positive impact both in terms of linear evaluation as well on the downstream. Also modifications of the classification head have slight but consistent positive effects.However, we observe more severe performance degradation upon reducing the number of hidden units of the LSTM modules from 512 to 256 (and 128), which confirms our assumption that the model capacity is of great importance. The most significant performance gain arises from finetuning in a two-step approach, where the head is finetuned first and the full model is only finetuned in a second step. Omitting discriminative learning rates in the final pretraining step lead to an identical mean performance as in the baseline case omitted, but the results are much less consistent across different runs as visible from a standard deviation that is almost double the size of the baseline value.

Finally, a comparison to \textit{CPC} applied to raw audio is in order. The original \textit{CPC} \cite{oord2018representation} applied to raw audio waveform data works on 10~kHz. The encoder uses strided convolutions and the encoded data therefore undergoes a downsampling by a factor of 160. Predicting 12 steps into the future then corresponds to a look-ahead interval of 0.192s. In our case, we work with much more coarsely sampled data at 100~Hz, but the encoded data undergoes no downsampling due to the use of a fully connected encoder. In this case, predicting 12 steps into the future corresponding to 0.12s, which lies in a similar order of magnitude as in speech. Using an encoder with strided convolutions lead to considerably worse performance that was already apparent for models trained in a supervised fashion.

\section{Performance improvements through self-supervised pretraining}
\renewcommand{\thefigure}{D\arabic{figure}}
\setcounter{figure}{0}
In this section, we provide additional details on the specific effects of self-supervised pretraining. In \Cref{fig:super_improv}, we aggregate the improvements according to superclasses revealing the largest improvements within the form and rhythm label categories. In \Cref{fig:improv}, we show the improvement through pretraining as a function of the supervised performance, which explicitly shows the negative correlation already mentioned in the main text.
\begin{figure}[t]
	\includegraphics[width=\columnwidth]{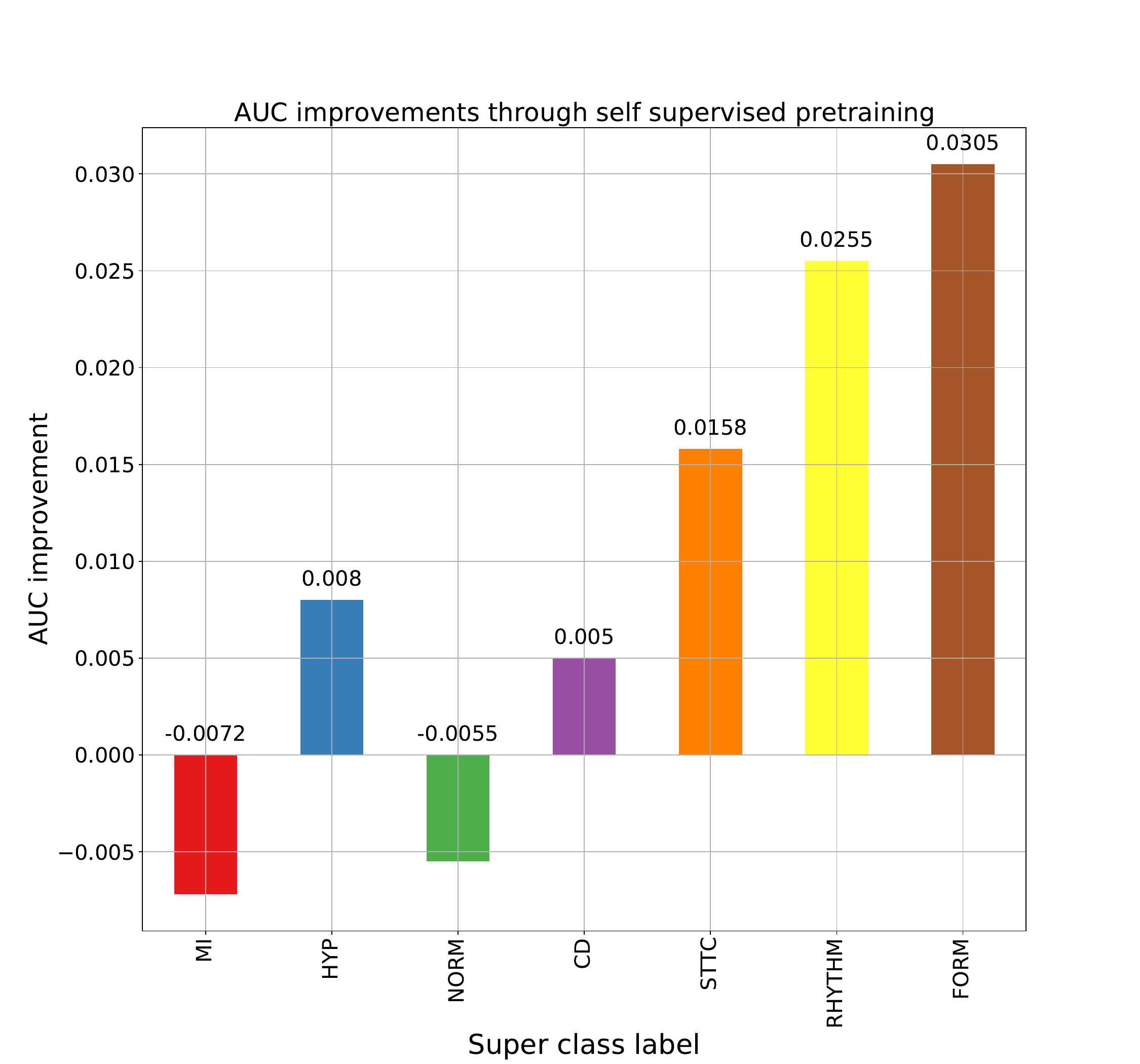}
	\caption{Improvement of classification of super level diagnosis, rhythm and form labels of the PTB-XL dataset of self-supervised training + finetuning in comparison to normal supervised training}
	\label{fig:super_improv}
\end{figure}

\begin{figure}[t]
	\includegraphics[width=\columnwidth]{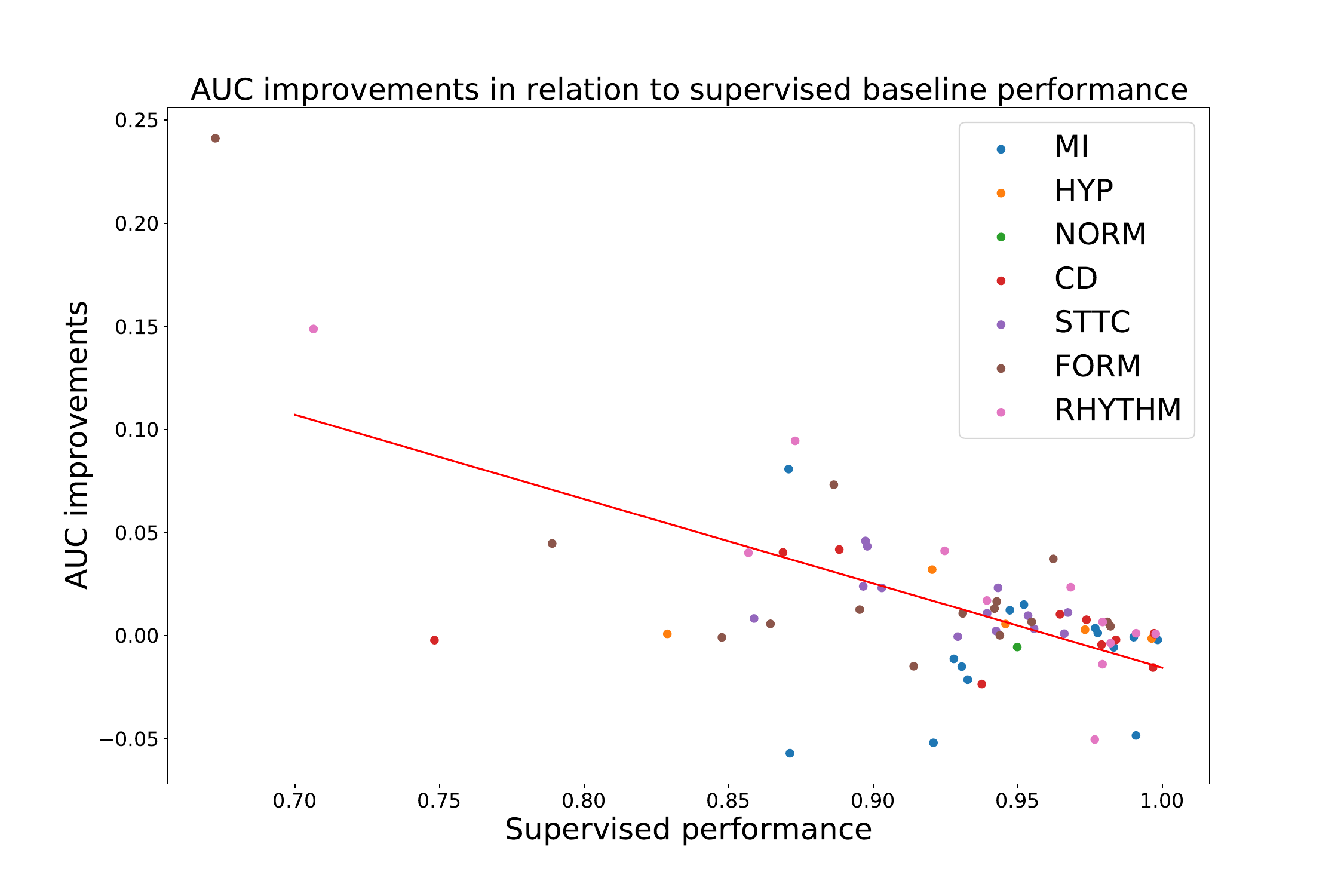}
	\caption{AUC improvement after self-supervised training as a function of supervised performance, where the red line represents linear regression of the data points.}
	\label{fig:improv}
\end{figure}

\section{Label distribution of CinC and PTB-XL}
\renewcommand{\thefigure}{E\arabic{figure}}
\setcounter{figure}{0}
\textit{PTB-XL} is a dataset that contains 21837 samples and is annotated with 71 labels at the finest level. The dataset used for the Computing in Cardiology Challenge 2020, here refered to as \textit{CinC2020}, was created by compiling five different datasets, including \textit{PTB-XL}. In order to create a unified set of labels, the orginal labels were mapped to SNOMED codes. This mapping process inevitably introduces ambiguities due to the fact that the original labels where assigned based on a different set of labels. Nevertheless the \textit{CinC2020} labels can be used to assess differences in data distributions of the different subsets of \textit{CinC2020}. Here, we are most interested in comparing the label distribution of \textit{PTB-XL} to \textit{CinC2020 w/o PTB-XL}, i.e. the subset of \textit{CinC2020} where all \textit{PTB-XL} records have been removed. Figure \ref{fig:label_distribution} compares the \textit{PTB-XL} dataset with the samples of the four remaining datasets in CinC, which sum up to 21256 signals. For the majority of pathologies, there are only a few samples in both datasets. For the pathologies that occur more frequently, large differences can be observed in many cases, indicating that the label distributions differ considerably.
\begin{figure}[t]
	\includegraphics[width=\columnwidth]{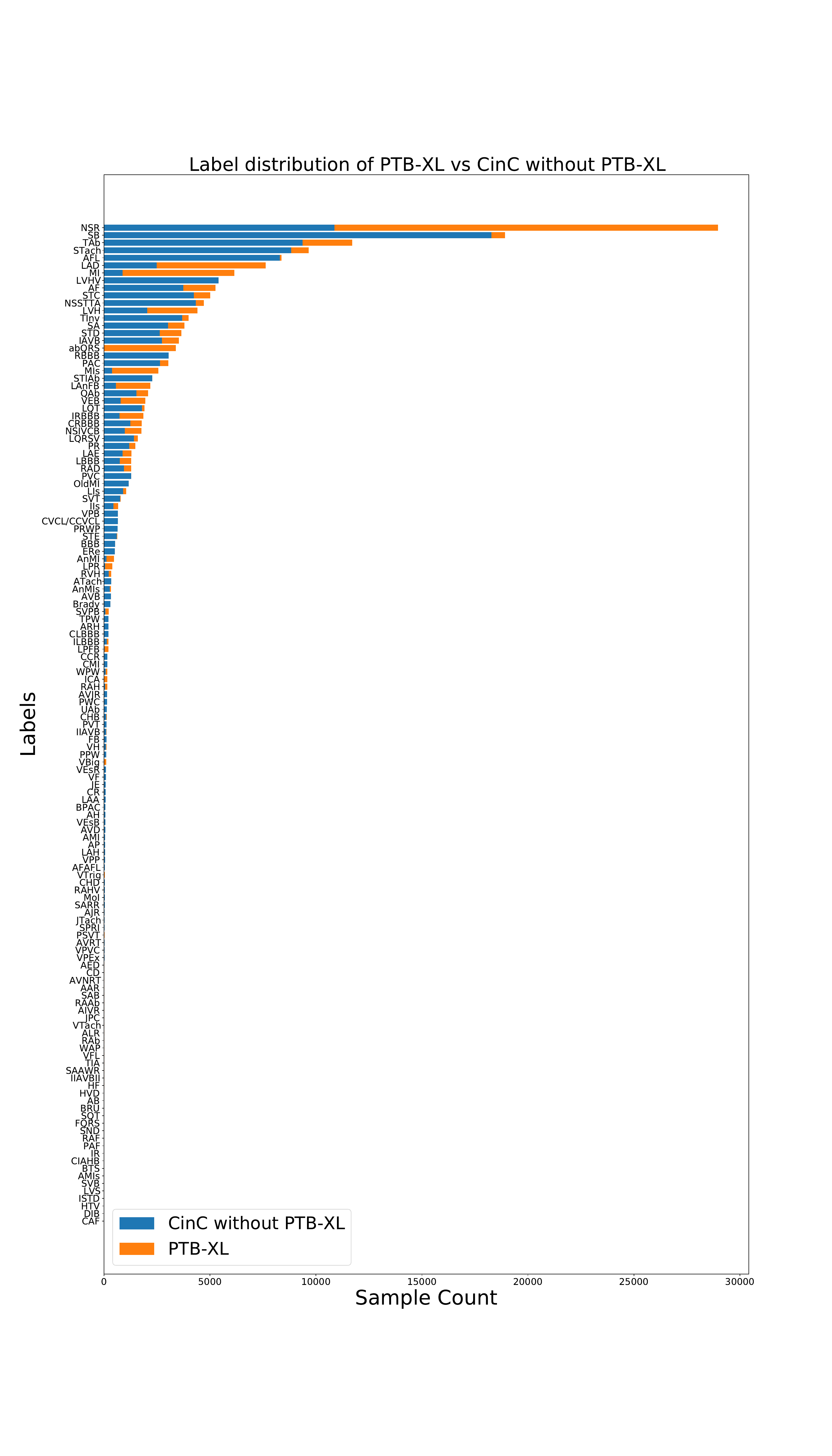}
	\caption{Comparison of the label distributions of \textit{PTB-XL} and \textit{CinC2020 w/o PTB-XL} in terms of labels provided within the \textit{CinC2020} dataset. Blue bars represent counts for \textit{CinC w/o PTB-XL}. Orange bars represent \textit{PTB-XL} counts.}

	\label{fig:label_distribution}
\end{figure}

        \end{appendices}

\end{document}